\documentclass[sigconf]{acmart}
\usepackage[svgnames]{xcolor}
\usepackage{xcolor}
\usepackage{graphicx}
\usepackage{subcaption}
\usepackage{enumitem}
\usepackage{enumerate}
\definecolor{purple}{HTML}{F3EFFE}
\definecolor{green}{HTML}{E0F7E5}
\definecolor{green+}{HTML}{CCFFE6}

\definecolor{purple+}{HTML}{D6C1FF}
\definecolor{blue}{HTML}{C9E3FF}
\definecolor{deepblue}{RGB}{0, 0, 139}

\usepackage[dvipsnames]{xcolor}
\usepackage[most]{tcolorbox}
\usepackage{amsmath}
\usepackage{algorithm}
\usepackage{algorithmic}
\usepackage{booktabs}
\usepackage{colortbl}
\usepackage{booktabs}
\usepackage{multirow}
\usepackage{graphicx}
\usepackage{makecell}
\usepackage{adjustbox}
\usepackage[dvipsnames]{xcolor}
\usepackage[normalem]{ulem}

\AtBeginDocument{%
  }

\setcopyright{acmlicensed}
\copyrightyear{2018}
\acmYear{2018}
\acmDOI{XXXXXXX.XXXXXXX}
\acmConference[Conference acronym 'XX]{Make sure to enter the correct
  conference title from your rights confirmation email}{June 03--05,
  2018}{Woodstock, NY}

\acmISBN{978-1-4503-XXXX-X/2018/06}

\newcommand\systemname{\textsc{MCTuner}}
\begin{document}
\begin{sloppypar}
\title{\systemname: Spatial Decomposition-Enhanced Database Tuning via LLM-Guided Exploration}

\author{Zihan Yan}
\orcid{0009-0001-1995-5312}
\affiliation{%
  \institution{University of Electronic Science and Technology of China}
  \city{Chengdu}
  \state{Sichuan}
  \country{China}
}
\email{yanzihan.yzh@std.uestc.edu.cn}

\author{Rui Xi}
\affiliation{%
  \institution{University of Electronic Science and Technology of China}
  \city{Chengdu}
  \state{Sichuan}
  \country{China}
}
\email{xirui@uestc.edu.cn}

\author{Mengshu Hou}
\affiliation{%
  \institution{University of Electronic Science and Technology of China}
  \city{Chengdu}
  \state{Sichuan}
  \country{China}
}
\email{mshou@uestc.edu.cn}
\thanks{Corresponding Author: Rui Xi (xirui@uestc.edu.cn)}

\renewcommand{\shortauthors}{Yan et al.}
\renewcommand{\shorttitle}{\systemname}
\renewcommand{\abstractname}{ABSTRACT}

\begin{abstract}
Database knob tuning is essential for optimizing the performance of modern database management systems, which often expose hundreds of knobs with continuous or categorical values. However, the large number of knobs and the vast configuration space make it difficult to identify optimal settings efficiently.
Although learning-based tuning has shown promise, existing approaches either ignore domain knowledge by relying solely on benchmark feedback or struggle to explore the high-dimensional knob space, resulting in high tuning costs and suboptimal performance.
To address these challenges, we propose \systemname, an adaptive knob tuning framework that minimizes exploration in ineffective regions of the configuration space. \systemname~employs a Mixture-of-Experts (MoE) mechanism with specialized LLMs to identify performance-critical knobs. 
In further, \systemname~introduces the first spatial decomposition algorithm that recursively partitions the space into hierarchical subspaces, on which Bayesian Optimization is performed to efficiently search for near-optimal configurations.
Evaluated on different benchmarks (OLAP, OLTP, and HTAP), \systemname~achieves up to 19.2\% performance gains and $1.4\times$ faster configuration discovery per iteration compared to state-of-the-art methods.
\end{abstract}

\received{20 February 2007}
\received[revised]{12 March 2009}
\received[accepted]{5 June 2009}

\maketitle

\section[INTRODUCTION]{INTRODUCTION}
Knob tuning in Database Management Systems (DBMS) is essential for enhancing system performance and ensuring efficient, stable operations. In practical applications, 
DBMS relies heavily on various knob configurations to achieve optimal performance, however, the complexity of tuning hundreds of interdependent parameters poses a persistent challenge \cite{zhang2024automatic,zhao2023automatic,zhang2021facilitating}. Improper configurations may lead to issues such as slow query responses and low resource utilization, severely impairing user experience and business operations. Consequently, automating and refining the knob tuning process has emerged as a pivotal research frontier, aiming to balance precision, adaptability, and computational overhead in dynamic workloads.

Traditional database knob tuning relies on heuristic methods, which are typically categorized into rule-based and search-based approaches. Rule-based methods are derived from manual tuning patterns, formulating tuning rules based on the experience of database administrators or database manuals \cite{sullivan2004using}. In contrast, search-based methods integrate various search techniques or employ sampling optimization strategies to partition the configuration space and explore subspaces to identify optimal knob settings \cite{ansel2014opentuner, zhu2017bestconfig}. Although heuristic methods are simple to implement, they often fail to achieve optimal configurations and are poorly equipped to handle dynamic changes in workloads and data.

To address these limitations, the academic community has introduced machine learning models, particularly Bayesian Optimization (BO) and Reinforcement Learning-based (RL-based) methods. BO utilizes surrogate models to approximate the objective function (database performance) and employs acquisition functions to balance exploration and exploitation, iteratively refining the tuning process \cite{duan2009tuning, van2017automatic, kunjir2020black, zhang2021restune, cereda2021cGPTuner, zhang2022towards, kanellis2022llamatune}. While this method is effective at identifying high-quality configurations, it is prone to getting trapped in local optima within large configuration spaces and is most suitable for continuous knobs, with limited adaptability to categorical ones. In contrast, RL frames the knob tuning problem within a reinforcement learning paradigm, where an agent interacts with the environment (the database) to learn the optimal tuning strategy \cite{zhang2019end, li2019qtune, wang2021udo, ge2021watuning, trummer2022db, cai2022hunter}. While RL excels at exploring high-dimensional spaces, it incurs significant tuning costs \cite{zhao2023automatic}.

Although the aforementioned methods have advanced database knob tuning and can achieve high-performance configurations, they still face three main challenges.

\textit{\textbf{C1: Exploration of invalid configurations wastes time and resources, hindering efficient optimization.}}
In databases, the default configurable space of knobs provided by DBMS vendors is vast. 
For example, in PostgreSQL, the \texttt{shared\_buffers} parameter is allocated within a configurable range constrained by the machine's RAM size. Specifically, when the RAM capacity is 128GB, the configurable range for \texttt{shared\_buffers} is [0, 128GB]. With the tuning unit for PostgreSQL set to 8KB, this range is further expressed as [0, 16,777,216]. 
Nevertheless, existing methods typically operate within the default configuration space without refining knob-specific domains or pruning ineffective regions \cite{cai2022hunter, cereda2021cGPTuner, li2019qtune, van2017automatic, zhang2019end, zhang2021restune}. Although recent learning-based techniques seek performance gains through complex modeling or exhaustive search strategies, these incur substantial computational overhead that impedes exploration efficiency and yields suboptimal configurations.
\textbf{A critical limitation is that they often overlook the fact that large portions of the configuration space are ineffective and contribute little to performance gains.} Consequently, the tuning process frequently spends many resources on exploring unproductive regions, leading to low efficiency and delayed convergence to optimal settings.
Fig.~\ref{ablation_space} illustrates the tuning performance for knobs, comparing the cases with and without the compression of configurable space. It shows that \systemname~integrated with LLM achieves optimal knob settings significantly faster than the approach without LLM. This highlights the importance of reducing the configuration space for knobs to enhance tuning efficiency.


\textit{\textbf{C2: The efficiency and interpretability of the knob selection remain inadequate.}}
Knob selection aims to optimize database performance and reduce tuning costs, but it faces several significant challenges. The relationship between knobs and performance is highly complex, and existing methods often inadequately address the intricate knob-performance relationships~\cite{debnath2008sard} and inter-knob correlations\cite{kanellis2020too, van2017automatic}. This oversimplification results in optimization blind spots while introducing inaccuracies. Gathering the necessary data to understand these relationships is further complicated due to the dynamic nature of workloads, as exhaustive configuration execution becomes prohibitively time-consuming. Crucially, current methods function as black boxes, offering limited transparency regarding how selections are made and how they align with specific client needs. 

\textit{\textbf{C3: Runtime samples scarcity under different workloads leads to prolonged adaptation costs.}}
Runtime samples serve as the foundation for learning-based methods, and the lack of high-quality samples often leads to inefficient database knob tuning. Generally, when a database encounters new workloads, conventional methods generate knob settings via sampling strategies or genetic algorithms \cite{van2017automatic,zhang2021restune,cai2022hunter}, then execute workloads on the newly configured database instance to collect runtime data. However, due to the inherent randomness, these stochastic methods frequently yield low-quality samples, necessitating resource-intensive data collection that prolongs adaptation and inflates computational overhead.

\underline{\textbf{Our Proposal.}}
To address the abovementioned challenges, we propose \textbf{\systemname}, a spatial decomposition-enhanced database tuning approach that leverages LLM-guided exploration to further enhance the optimization process. 

\textit{To address challenge C1, we propose a spatial decomposition algorithm based on Monte Carlo Tree Search (MCTS) to recursively reduce the configurable space.}
Inspired by LA-MCTS \cite{wang2020learning}, our approach decomposes vast, high-dimensional, and heterogeneous configurable space and selects optimal knobs within well-performing subspaces, significantly improving the tuning efficiency. 
To enable progressive compression of the configuration space, our spatial decomposition algorithm adaptively shrinks the knob search space during tuning. Based on the available sample size, it selects appropriate clustering techniques to partition the space into high- and low-performance regions. The overall tuning workflow integrates Bayesian Optimization (BO) with this decomposition strategy by recursively partitioning the space via a tree structure. During each iteration, promising regions are selected using the Upper Confidence Bound (UCB) criterion, within which BO is applied to sample and optimize configurations.


\textit{Concerning challenge C2, we propose a Mixture-of-Experts (MoE) mechanism to manage tuning knowledge from multiple sources for knob selection, leveraging the powerful language understanding and generation capabilities of LLMs.} By collecting knowledge from diverse sources- including database manuals, web information, and LLM-generated content-and rigorously validating this information, we establish a comprehensive and accurate foundation for knob selection. The MoE mechanism incorporates seven domain experts tasked with selecting key knobs based on multi-source knowledge and user needs. This process encompasses \textit{knob classification}, \textit{weight assignment}, \textit{expert evaluation}, and \textit{weighted ranking and selection}, ultimately enhancing the accuracy and relevance of the selection. \textit{Regarding the challenge C3, we initiate the spatial decomposition algorithm with a limited number of samples, effectively narrowing the configurable space.} This approach significantly reduces the likelihood of generating ``bad'' configurations, thereby ensuring the efficiency of database knob tuning. 


In our evaluation, we compare \systemname~with seven state-of-the-art tuning methods across eight representative PostgreSQL benchmarks, covering a diverse range of workloads including OLAP, OLTP, and HTAP. \systemname~identifies high-quality configurations 1.4× faster per iteration and achieves up to 19.2\% performance improvement, measured by increased throughput or reduced latency, over the best-performing baselines. The MoE-based knob selection also outperforms other selection methods in both tuning efficiency and final performance. Furthermore, \systemname~adapts quickly to workload drift and exhibits strong transferability of learned configurations across workloads. These results collectively demonstrate the effectiveness and robustness of \systemname~for automatic database knob tuning in diverse and dynamic scenarios.


In summary, we make the following contributions:

\begin{itemize}[leftmargin=*]
\item We introduce a Mixture-of-Experts mechanism that leverages multi-source knowledge through seven domain-specific experts, enhancing selection accuracy and interpretability.
\item We propose a tuning strategy that recursively partitions high-dimensional configuration spaces via Monte Carlo Tree Search, reducing the search space and enhancing the likelihood of discovering optimal configurations.
\item By combining these two components, we design \systemname, the first spatial decomposition-enhanced database tuning framework with LLM-guided exploration.




\item We conduct extensive experiments across diverse benchmarks, performance metrics, and dynamic scenarios to demonstrate the effectiveness of \systemname.
\end{itemize}



\section{Related Work}
\subsection{Database Knob Tuning}
Modern DBMSs provide hundreds of configurable knobs designed to optimize performance and resource utilization \cite{van2021inquiry}. The effective management of these knobs is critical for improving database efficiency. Existing works generally encompass two primary approaches: heuristic-based methods and learning-based methods. Notably, among the learning-based techniques, BO and RL have demonstrated considerable success. To effectively tackle the workload variations, transfer learning has also been adopted~\cite{zhao2023automatic}.

\subsubsection{Heuristic-based methods.}
Heuristic methods reduce the tuning space using predefined rules or search strategies, which are classified into rule-based and search-based approaches. Rule-based methods depend on manually crafted tuning rules \cite{dageville2002sql}, while search-based methods optimize based on assumed distributions of the knob configuration space. For instance, OpenTuner \cite{ansel2014opentuner} integrates multiple search techniques, and BestConfig \cite{zhu2017bestconfig} combines sampling with local search for improved efficiency. However, both approaches are limited by the quality of initial samples and sampling costs. Although simple to implement and quick to converge, heuristic methods rely heavily on manual intervention and often struggle with search efficiency, limiting their ability to achieve optimal performance \cite{zhao2023automatic}.

\subsubsection{Learning-based methods.}
Machine learning has revolutionized database tuning, with BO and RL being the primary methods. BO builds surrogate models to balance exploration and exploitation, while RL learns tuning policies through interactive optimization. Despite their advantages, both methods face scalability and workload adaptation challenges.

\textbf{BO} uses surrogate models to link knob configurations with performance, updating them with prior knowledge and new observations to improve efficiency and accuracy. Common models include Gaussian Processes (GP) in iTuned \cite{duan2009tuning}, random forest in LlamaTune \cite{kanellis2022llamatune} and GPTuner \cite{lao2023GPTuner}, as well as Tree-structured Parzen Estimators (TPE) \cite{zhang2021facilitating} and Bayesian neural networks \cite{graves2011practical,hernandez2015probabilistic}.

Recent studies focus on enhancing surrogate model expressiveness by incorporating system-level features. OnlineTune \cite{zhang2022towards} combines the GP-UCB acquisition function and utilizes query features to improve model performance; CGPTuner \cite{cereda2021cGPTuner} employs GP-Hedge \cite{hoffman2011portfolio} to dynamically select acquisition functions and integrates workload-specific system attributes; RelM \cite{kunjir2020black} addresses memory control in containerized environments; ResTune \cite{zhang2021restune} incorporates CPU, memory, and I/O utilization to optimize tuning. However, BO's scalability is limited by its computational complexity, especially in high-dimensional spaces, with performance dropping due to its $O(n^3)$ complexity \cite{kanellis2020too}. To address this, researchers are exploring more efficient surrogate models.

\textbf{RL} autonomously optimizes database knobs through interactive learning and demonstrates strong exploration capabilities in unknown or large-scale configuration spaces \cite{ge2021watuning,li2019qtune,zhang2019end}. Compared to BO, RL relies on the Deep Deterministic Policy Gradient (DDPG) \cite{silver2014deterministic} to handle continuous action spaces and employs an actor-critic structure for knob tuning. CDBTune \cite{zhang2019end} utilizes DDPG to generate configurations but struggles with adaptability under dynamic workloads. QTune \cite{li2019qtune} incorporates query features to enhance tuning performance, while WATuning \cite{ge2021watuning} leverages pretrained models and attention mechanisms to improve generalization. UDO \cite{wang2021udo} reduces restart costs, and HUNTER \cite{cai2022hunter} and DB-BERT \cite{trummer2022db} accelerate RL training through rule extraction or sample generation. Although these approaches enhance tuning efficiency, obtaining high-quality rules remains costly. Compared to BO, RL requires less prior knowledge but tends to face instability issues in large-scale knob spaces. Researchers are actively exploring more efficient training strategies and sample utilization mechanisms to improve the applicability.

\subsubsection{knowledge transfer methods.}
Knowledge transfer methods leverage prior experience to enhance the tuning efficiency of new workloads, primarily through workload mapping, model pretraining, and model ensemble strategies. Workload mapping methods (e.g., OtterTune \cite{van2017automatic} and CGPTuner \cite{cereda2021cGPTuner}) initialize models based on historical workload similarity, accelerating convergence. Model pretraining approaches (e.g., QTune \cite{li2019qtune}, WATuning \cite{ge2021watuning}, and OpAdvisor \cite{zhang2023efficient}) integrate detailed workload features to improve tuning performance but may suffer from overfitting. Model ensemble techniques (e.g., ResTune \cite{zhang2021restune}) combine multiple models to adapt to diverse workloads, effectively mitigating cold-start issues, though balancing contributions from different models remains challenging. Additionally, knob selection strategies (e.g., GPTuner \cite{lao2023GPTuner} and OpAdvisor \cite{zhang2023efficient}) refine the configurable space to improve tuning efficiency. While these methods enhance tuning performance, database tuners still necessitate adaptation to effectively address new workloads. Consequently, minimizing the number of tuning iterations required for these novel workloads has emerged as a primary focus of research in this domain.
\subsection{LLM for Knob Tuning}
Recently, LLMs have been increasingly applied to key areas of database research, including knob tuning \cite{trummer2022db, lao2023GPTuner, huang2025e2etuneendtoendknobtuning, giannakouris2025lambda, li2024large}, diagnosis \cite{zhou2023d, singh2024panda, ouyang2025rcrank, giannakouris2024dbg, xiu2024query}, text-to-SQL \cite{li2023resdsql, li2024codes, pourreza2023din, li2024dawn, ren2024purple, sivasubramaniam2024sm3}, query optimization \cite{trummer2022codexdb, li2024llm, akioyamen2024unreasonable, sun2024r}, and index recommendation \cite{zhao2025llmidxadvis, zhou2024breaking}. These studies highlight LLMs’ potential to advance intelligent DBMS management.
In database knob tuning, LLMs optimize configurations to enhance system performance and automate tuning. DB-BERT \cite{trummer2022db} and GPTuner \cite{lao2023GPTuner} use knowledge from forums and manuals to refine the configurable space, improving tuning efficiency. Specifically, DB-BERT combines reinforcement learning and benchmarking feedback for optimization, while GPTuner unifies structured knowledge with Bayesian Optimization for efficient selection. E2ETune \cite{huang2025e2etuneendtoendknobtuning} fine-tunes models to directly recommend knob configurations based on workload characteristics, reducing iterations compared to traditional methods. $\lambda$-Tune \cite{giannakouris2025lambda} uses prompt engineering and dynamic programming for OLAP workload tuning. 
These advancements show LLMs improve automation and open new avenues in database configuration.

However, existing LLM-based methods suffer from several shortcomings. Primarily, they rely on a simplistic, one-step approach for knob selection that does not incorporate multiple processing stages and treats all knobs equally. Moreover, the knob tuning process functions as a black box with insufficient analysis of how individual knobs affect specific database functions.

\begin{figure*}
\centering
\includegraphics[width=1\linewidth]{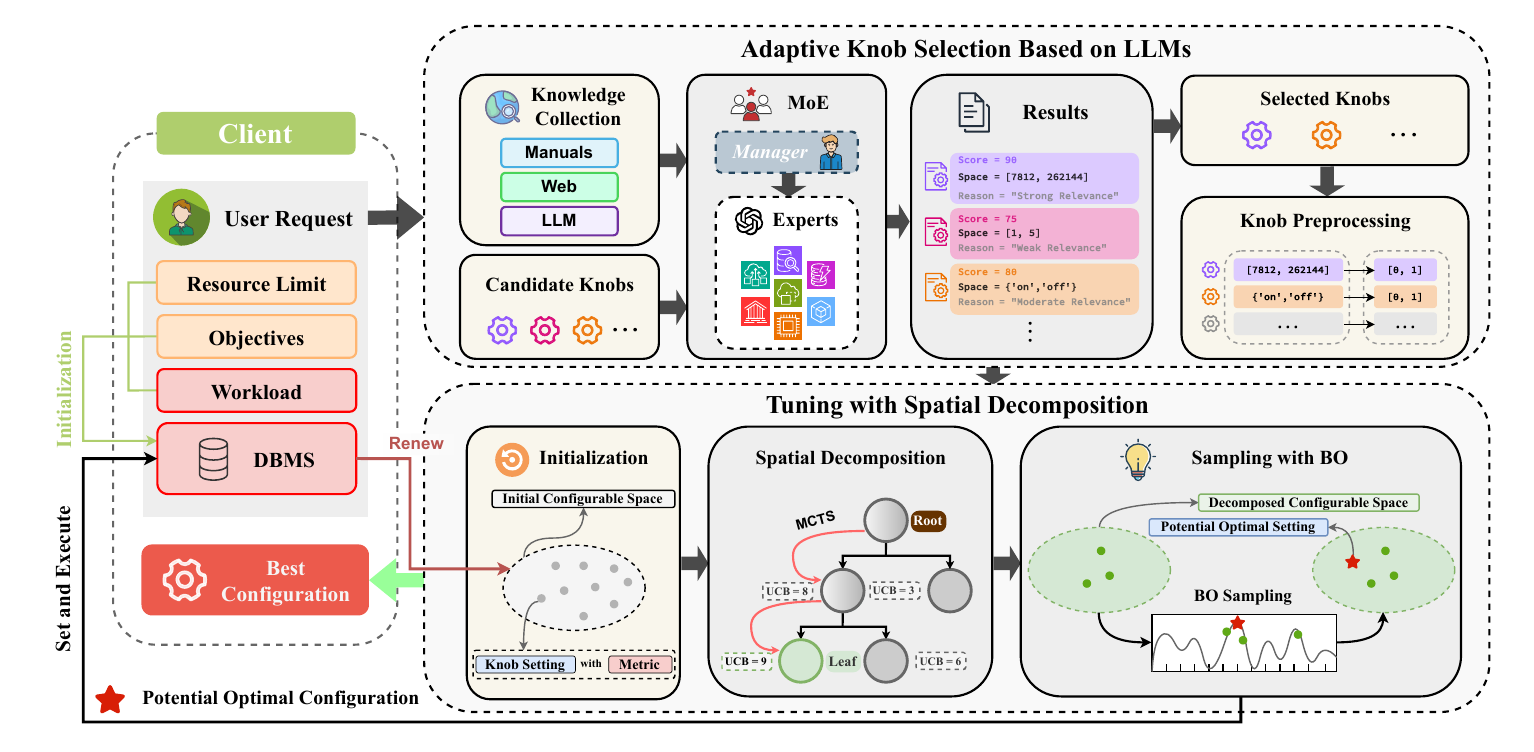}
\caption[Overview of the system]{Overview of \systemname.}
\vspace{-0.5em}
\label{FIG_overview}
\vspace{-0.5em}
\end{figure*}

\section{PROBLEM DEFINITION}

In database management systems, configurable tuning parameters (knobs) govern critical system attributes including memory allocation, connection concurrency, and query optimization strategies. Formally, let $K = \{k_1, k_2, \ldots, k_n\}$ denote the set of tunable knobs, where each parameter $k_i$ accepts values from either a continuous domain or or categorical set, defining its configuration space $R_i$. The joint configuration space $R = R_1 \times R_2 \times \ldots \times R_n$ represents the combinatorial multidimensional space of all possible parameter combinations. And each specific configuration is represented by a set of knob values $\mathbf{r} = (r_1, r_2, \ldots, r_n) \in R$, $r_i$ is the value of the knob $k_i$.

Given a database instance DB and workload $W = \{q_1, \ldots, q_m\}$ operational queries, knob tuning aims to identify the optimal configuration vector $\mathbf{r}^\star = (r_1^\star, r_2^\star, \ldots, r_n^\star) \in R$ that extremizes target performance metrics $M$, such as minimizing query latency or maximizing transaction throughput under constrained resources.

\section{MCTUNER OVERVIEW}
The overview of \systemname~is illustrated in Fig.~\ref{FIG_overview}, which comprises two main steps.

\textbf{Step 1: Adaptive Knob Selection Based on LLMs (Section 5).} 
In the first step, knob selection is formulated as a dimensionality and range reduction problem, aiming to eliminate ineffective regions within the configuration space and enhance tuning efficiency.
\systemname~aggregates domain knowledge from database manuals, web resources, and LLM-generated insights. It employs a MoE framework in which specialized LLM-based experts dynamically assess knob importance and collaboratively select a performance-critical subset.
To enable efficient joint optimization, categorical knobs are further encoded into continuous representations.

\textbf{Step 2: Tuning Based on Spatial Decomposition (Section 6).} In this step, the configurable space is progressively decomposed to enable more focused and efficient exploration. 
Optimization objectives are first weighted based on user requirements and LLM-inferred priorities.
\systemname~then applies a spatial decomposition algorithm that recursively partitions the space using a search tree guided by MCTS. To adaptively refine subspaces, clustering is performed based on sampling density, and a soft-margin SVM filters out unpromising regions. Bayesian Optimization is applied within selected leaf nodes to identify high-quality configurations. The search tree is dynamically reconstructed across iterations to incrementally converge toward the optimal knob setting for the target environment.

\systemname~combines static and dynamic space reduction to enable efficient and targeted tuning. LLM-based knob selection filters out irrelevant knobs and shrinks value ranges, while spatial decomposition progressively refines the search to focus on high-potential regions. This two-stage design mitigates ineffective exploration and enhances the probability of identifying optimal configurations.

\section{ADAPTIVE KNOB SELECTION WITH LLMS}\label{adaptive}
\subsection{Multi-Source Knowledge Collection}
\begin{figure}
\centering
\includegraphics[width=1\linewidth]{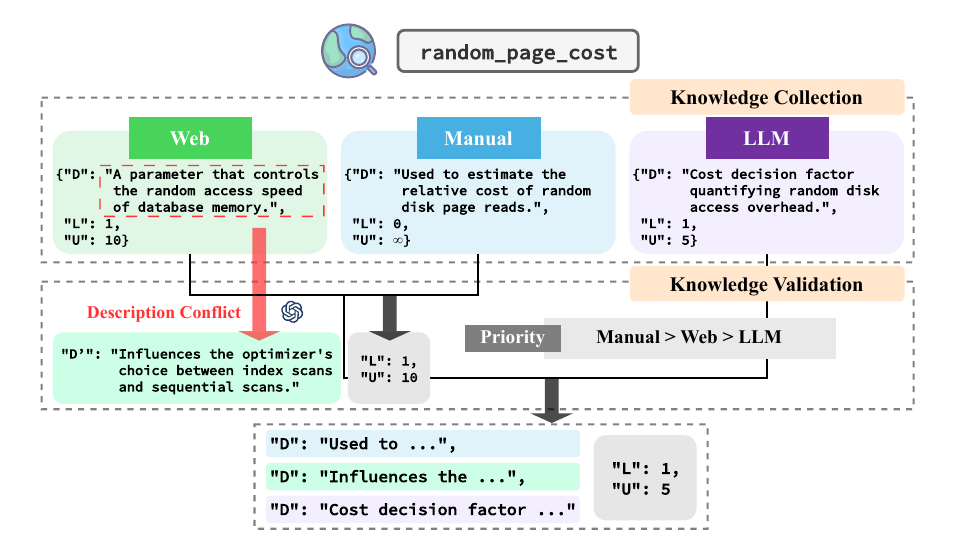}
\caption{The workflow for collecting multi-source knowledge of the knob \texttt{random\_page\_cost}.}
\label{FIG_1}
\vspace{-1em}
\end{figure}

To address the dimensionality reduction challenge, \systemname~avoids collecting execution data during tuning for efficiency consideration, and instead leverages the reasoning capabilities of LLMs to guide the reduction process.
This process is driven by multi-source heterogeneous knowledge, with LLMs playing a central role in its extraction and interpretation \cite{zhang2023igniting}.
Inspired by GPTuner \cite{lao2023GPTuner}, we divide the knowledge processing workflow into two main steps: \textbf{knowledge collection}, and \textbf{knowledge validation}. This structure integrates multi-source knowledge and provides strong support for subsequent knob selection, as illustrated in the Fig.~\ref{FIG_1}.

In the \textbf{knowledge collection} phase, we gather descriptive information about the knobs and their recommended configurable spaces, represented as $\mathcal{K} = \{(D_i, R_i) \mid i \in \{W, LLM, M\} \}$, where $D_i$ denotes the knob description and $R_i$ is the corresponding recommended configuration space, both derived from source $i$ (i.e., webpage information ($W$), large language models ($LLM$), or official DBMS manuals ($M$)). For continuous and integer knobs, $R$ is defined by a lower bound $L$ and an upper bound $U$. In contrast, categorical knobs take values from a small, finite set (typically no more than 10 options in PostgreSQL). Since categorical knobs have limited variability, we focus the following discussion on continuous and integer knobs, which are characterized by the interval $[L, U]$.

To construct $\mathcal{K}$, we combine information from manuals, web pages, and LLMs. Each source contributes complementary insights: manuals provide authoritative definitions, web scraping supplements missing details, and LLMs generate tuning-specific knowledge. This multi-source strategy ensures coverage and completeness by addressing the limitations of individual sources.
For example, if the official PostgreSQL manual lacks specific recommendations for a knob (e.g., \texttt{random\_page\_cost}), LLM can provide supplementary insights.

\begin{center}
\begin{tcolorbox}[colback=purple,
                  colframe=black,
                  width=1\linewidth,
                  boxrule=0.5pt,
                  arc=3mm, auto outer arc,
                 ]
LLM = \{"$D$": "\texttt{random\_page\_cost} estimates the cost of random disk page reads, guiding query plan selection. It should be tuned based on disk I/O, workload patterns, and caching conditions.","$L$": "1","$U$": "10"\}
\end{tcolorbox}
\end{center}

To ensure accurate knob-related knowledge, we must validate and filter tuning information in the \textbf{knowledge validation} phase. This process guarantees reliable and reasonable tuning recommendations based on a comprehensive analysis of multiple sources. For each knob $k$, we first validate the knowledge $\mathcal{K}_k$ internally, using LLM’s contextual learning to filter out noise. A key step is to check if the descriptive information is extreme and take the intersection of the configurable space from multiple sources, i.e., $L_k = max(L_{M_k}, L_{LLM_k}, L_{M_k})$, $U_k = min(U_{M_k}, U_{LLM_k}, U_{M_k})$ (via LLM).

When summarizing descriptive information, we prioritize sources in the following order: official manuals ($D_{M_k}$) > web content ($D_{W_k}$) > LLM ($D_{LLM_k}$). This priority ensures that tuning recommendations are mainly based on official documentation. In cases where the official source lacks clarity, we turn to online communities, and subsequently, we consider the generative capabilities of LLMs. This aims to minimize the risk of inaccurate recommendations.

In managing information related to configurable spaces, we treat all sources with equal importance, placing particular emphasis on knowledge derived from web content and LLMs. This approach arises from the observation that the configurable ranges suggested in official manuals can often be overly broad.
For instance, the manual for \texttt{random\_page\_cost} suggests a range of $[0, 1.79769 \times 10^{308} (\infty)]$. However, setting \texttt{random\_page\_cost} to $4$ typically yields better database performance. Therefore, relying solely on the configurable spaces provided in the official manual to guide further tuning is not advisable.

After validating and resolving any conflicts, we generate a knowledge summary for each knob $k$: $\mathcal{S}_k=(D_{M_k}, D_{W_k}, D_{LLM_k}, [L_k, U_k])$, which offers high-quality knowledge support for the subsequent knob selection process.

\subsection{Adaptive Knob Selection with MoE}

Since different knobs exert varying degrees of influence on system performance, thereby making the selection of critical knobs essential for effective optimization. However, existing knob selection methods often fail to dynamically adapt to workload variations and lack interpretable analysis of the specific database functions controlled by each knob, resulting in a black-box selection process.

To address these limitations, we introduce the Mixture-of-Experts (MoE) mechanism, which optimizes knob selection by integrating seven domain-specific LLM experts, each specializing in handling one category of knobs, managed by the expert \textit{Manager}. Specifically, these experts specialize in \textit{Access Control}, \textit{Query Optimization}, \textit{Query Execution}, \textit{Background Processes}, \textit{CPU}, \textit{Memory}, and \textit{Disk} \cite{zhao2023automatic}. They analyze multi-source knowledge alongside user-defined requirements to enhance the selection process. Furthermore, to avoid excessively long outputs during interactions with LLMs using traditional Chain-of-Thought (CoT), we employ Chain of Draft (CoD) \cite{xu2025chain} to retain only the key information.

The MoE mechanism follows a collaborative expert evaluation approach to identify the most critical knobs. The summarized knowledge ($\{\mathcal{S}_k\}$) serves as auxiliary information for experts, guiding a three-step process:

\begin{itemize}
\item[Step 1] \textbf{Knob Classification and Weight Allocation} – Classifies knobs and assigns appropriate weights.
\item[Step 2] \textbf{Expert Evaluation} – Domain experts assess the importance of the assigned knobs and provide concise justifications.
\item[Step 3] \textbf{Weighted Ranking and Selection} – Aggregates expert evaluations to rank and filter the most impactful knobs for optimization. 
\end{itemize}

\begin{figure*}
\centering
\includegraphics[width=1\linewidth]{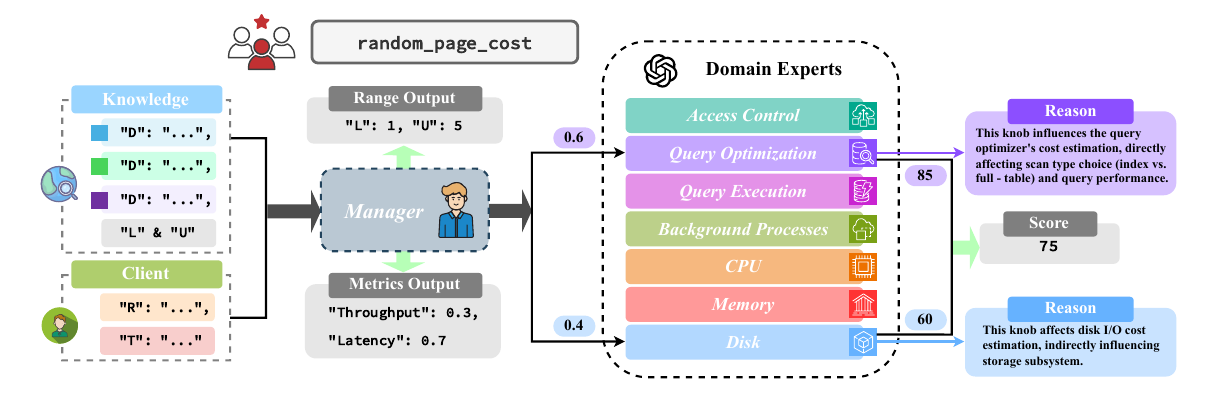}
\caption{The workflow of MoE.}
\label{FIG_2}
\end{figure*}

This structured selection process ensures that the chosen knobs significantly improve database tuning efficiency. The overall workflow is illustrated in Fig.~\ref{FIG_2}.

\subsubsection{Knob Classification and Weight Allocation}
When a candidate knob ($k$) is input into the system, its heterogeneous information $\mathcal{S}_k$ is first extracted and combined with the user requirements ($\mathcal{R}$) and task description ($\mathcal{T}$) to help the LLM understand the optimization objective.

\begin{itemize}[leftmargin=*]
\item $\mathcal{R}$ specifies the user's tuning requirements, which are typically used to determine the key metrics for subsequent tuning.
\item $\mathcal{T}$ describes the target database for tuning (e.g., PostgreSQL), workload (e.g., TPC-H, TPC-C), and hardware configuration. \end{itemize}

Hardware configuration is crucial because the upper and lower bounds collected during the \textbf{Knowledge Collection} step may be expressed as relative values. Since database knob settings cannot use relative values directly, we must compute precise values based on the specific hardware. For example, if the bounds for \texttt{effective\_cache\_size} are "L": "50\% of RAM", "U": "75\% of RAM", only with user input such as RAM = 16GB in $\mathcal{T}$ can we compute the actual range as [8GB, 12GB].

Subsequently, \textcolor{deepblue}{$(\mathcal{S}_k, \mathcal{R}, \mathcal{T})$} is passed to the \textit{Manager} to determine the configurable space of the knobs (as described above) and to classify them. Additionally, the \textit{Manager} assigns weights to each category based on task relevance, ensuring that the total weight sum equals 1. For knob categories that are irrelevant to the current task, no weight is assigned, thereby reducing redundant computations.

\begin{figure}
\centering
\includegraphics[width=1\linewidth]{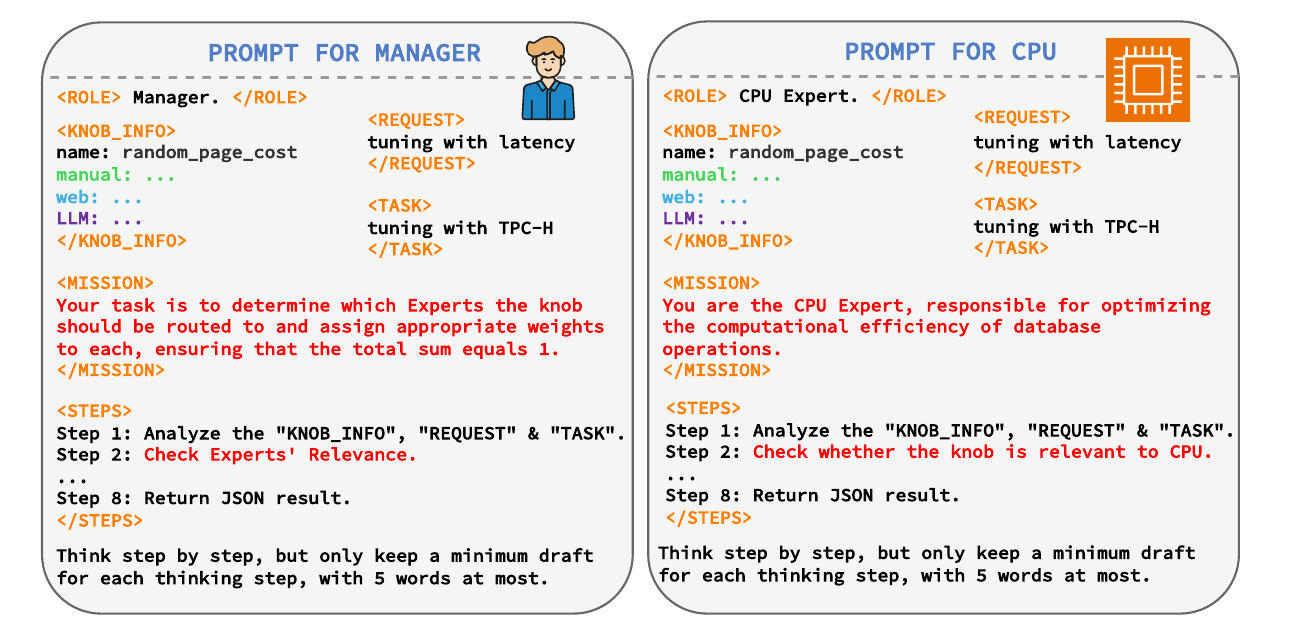}
\caption{Prompt Templates of MoE.}
\vspace{-1em}
\label{FIG_3}
\end{figure}

For example, in an optimization task focused on query performance, where \texttt{random\_page\_cost} is a candidate knob, the \textit{Manager} classifies it under \textit{Query Optimization} and \textit{Disk}, assigning respective weights (e.g., 0.6 and 0.4). Since \texttt{random\_page\_cost} primarily affects disk access rather than memory management, the \textit{Memory} is not assigned any weight, thus minimizing computational overhead. The prompt for this step is illustrated on the left side of Fig.~\ref{FIG_3}.

\subsubsection{Expert Evaluation}
After completing knob classification, the administrator assigns the knob information to 1 to 7 domain experts for independent evaluation. Each expert scores the knob based on its significance in their respective field, using a scale of 1 to 100, where 1 indicates minimal impact on system optimization, and 100 denotes critical importance. Additionally, experts must provide a brief justification for their score, explaining their assessment.

For instance, consider the knob \texttt{random\_page\_cost}:

\begin{center}
\begin{tcolorbox}[colback=purple+,
                  colframe=black,
                  width=1\linewidth,
                  boxrule=0.5pt,
                  arc=3mm, auto outer arc,
                 ]
\textit{Query Optimization}: \{"score": 85, "reason": "This knob influences the query optimizer’s cost estimation, directly affecting the choice between index scans and full table scans, which impacts query execution performance."\}
\end{tcolorbox}
\end{center}

By involving multiple experts for independent evaluations, the MoE mechanism captures diverse perspectives, reducing bias and ensuring a more comprehensive, accurate assessment. The prompt for this step is illustrated on the right side of Fig.~\ref{FIG_3}.

\subsubsection{Weighted Ranking and Selection}
After the expert evaluation is complete, we combine the category weights assigned by the \textit{Manager} with the expert scores to compute the final importance score for each knob and rank them accordingly. The calculation is given by: $S_{\text{final}} = \sum_{i=1}^{N} W_i \times S_i$, 
where \( S_i \) represents the importance score assigned by the \( i \)-th expert, \( W_i \) denotes the weight assigned to that expert, and \( S_{\text{final}} \) is the final importance score of the knob.

For example, for \texttt{random\_page\_cost}, the \textit{Query Optimization} expert assigns a score of 85 with a weight of 0.6, while the \textit{Disk} expert assigns a score of 60 with a weight of 0.4. The final importance score is calculated as: 
$S_{\text{final}} = (0.6 \times 85) + (0.4 \times 60) = 75$.

Once the importance scores for all knobs are computed, the system selects the top $N$ knobs with the highest scores as the most critical candidates for optimization. Therefore, \systemname~performs dimensionality reduction on the knob set as the initial step of configuration space compression.

\section{SPATIAL DECOMPOSITION BASED KNOB TUNING}
\begin{figure*}
\centering
\includegraphics[width=1\linewidth]{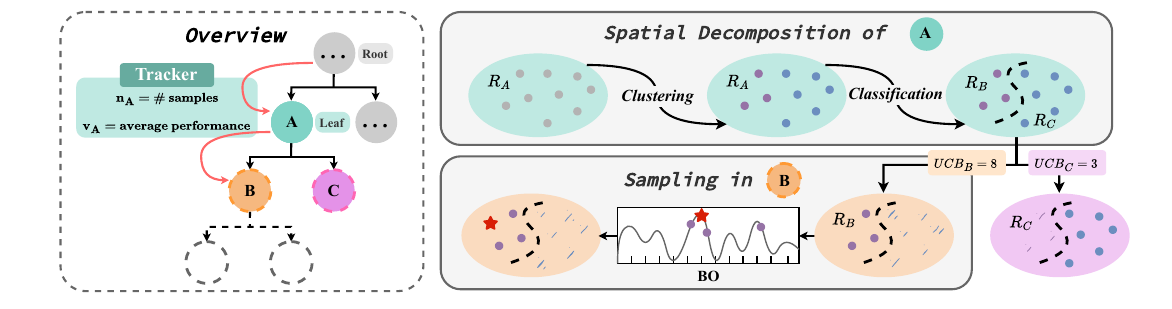}
\caption{Spatial Decomposition Based Knob Tuning.}
\label{FIG_4}
\end{figure*}
\subsection{Tuning Purpose}
In database knob tuning, selecting appropriate optimization metrics, typically throughput and latency, is essential. \systemname~leverages LLMs to recommend weights for these metrics based on $\mathcal{R}$ and $\mathcal{T}$. During the \textit{Knob Classification and Weight Allocation} step, the \textit{Manager} assigns weights $w_1$ and $w_2$ to throughput and latency, respectively; this process is performed once, as illustrated in Fig.\ref{FIG_2}. The optimization objective focuses on relative improvements ($p$) rather than absolute performance values:
\begin{equation}
   p(\mathbf{r}) =  w_1\frac{tps-tps_{\text{default}}}{tps_{\text{default}}}+w_2\frac{lat_{\text{default}}-lat}{lat_{\text{default}}}
\end{equation}
where $tps$ and $lat$ denote the throughput and latency under the current knob configuration $\mathbf{r}$, while $tps_{\text{default}}$ and $lat_{\text{default}}$ refer to the corresponding metrics under the default configuration.

\subsection{Spatial Decomposition Algorithm}

After the MoE mechanism statically compresses the configuration space, the overall space is significantly reduced. However, due to the high dimensionality and the inherently conservative nature of LLM-generated recommendations, the search space remains large. To address this, \systemname~further refines the space dynamically during tuning by applying a spatial decomposition algorithm. By partitioning the space into smaller subspaces, the algorithm focuses the optimization toward the most promising regions, reducing tuning complexity and increasing the likelihood of discovering optimal configurations.
This strategy improves the precision of knob setting selection while alleviating the computational cost of exhaustively exploring the full space.

\begin{algorithm}[t]
\caption{\textsc{Spatial\_Decomposition}}
\label{ALG_1}
\renewcommand{\algorithmicrequire}{\textbf{Input:}}
\renewcommand{\algorithmicensure}{\textbf{Output:}}
\begin{algorithmic}[1]
\REQUIRE Dataset $D_t$, region $R_A$
\ENSURE Two subregions $R_B$, $R_C$

\STATE Extract $D_A = \{(\mathbf{r}_i, p(\mathbf{r}_i)) \in D_t \mid \mathbf{r}_i \in R_A\}$
\STATE Compute $n_A \gets |D_A|$, $v_A \gets \text{mean}(p(\mathbf{r}_i))$
\IF{$n_A \leq \tau$}
    \STATE Apply spectral clustering based on cosine similarity
\ELSE
    \STATE Apply Kernel PCA followed by K-medoids clustering
\ENDIF
\STATE Assign cluster labels $\{y_i\}$ to all samples
\STATE Identify cluster with higher average $p(\mathbf{r}_i)$ as $R_B$, else $R_C$
\RETURN $R_B$, $R_C$
\end{algorithmic}
\end{algorithm}

In each iteration $t$, \systemname~collects an evaluation dataset $D_t = \{(\mathbf{r}_i, p(\mathbf{r}_i))\}$, where $\mathbf{r}_i$ denotes a knob configuration and $p(\mathbf{r}_i)$ its observed performance.

A tree node (e.g., node A in Fig.\ref{FIG_4}) corresponds to a subregion $R_A \subseteq R$, and the subset $D_A = \{(\mathbf{r}_i, p(\mathbf{r}_i)) \in D_t \mid \mathbf{r}_i \in R_A\}$ contains all samples located within this region. Each node tracks two key statistics to guide the search process:
\begin{itemize}[leftmargin=*]
\item $n_A$: The number of samples in node A, computed as $n_A = |D_A|$
\item $v_A$: The average performance of node A, computed as: $v_A = \text{mean}(p(\mathbf{r}_i)), \forall (\mathbf{r}_i, p(\mathbf{r}_i)) \in D_A$
\end{itemize}

\systemname~~progressively narrows the configuration space using a recursive spatial decomposition strategy, initialized from a root node built over the full dataset $D_t$. At the beginning, this root node also serves as the only leaf node. For each current leaf node (e.g., node $A$ in Fig.~\ref{FIG_4}), clustering and classification procedures are applied as described in Alg.~\ref{ALG_1} to divide its associated subregion $R_A$ into two non-overlapping parts: a high-performance region $R_B$ and a low-performance region $R_C$. By convention, the high-performance region is assigned to the left child.

The division is based on learned decision boundaries and reflects differences in performance across clusters, enabling the system to concentrate future sampling on more promising areas. As the tree grows, each recursive split further isolates high-value regions. For instance, node $B$ in Fig.~\ref{FIG_4} represents a refined subregion resulting from repeated decomposition, where prior evaluations indicate strong performance potential.

This strategy effectively prunes low-reward areas from the search space and improves tuning efficiency by prioritizing exploration of subspaces most likely to yield optimal configurations. In the following subsections, we will provide a detailed explanation of the clustering and classification methods used by \systemname~during the spatial decomposition process, along with their specific implementation steps.

\subsubsection{Clustering Method}
Tuning database knobs typically requires re-collecting samples as workloads change. Each sample adjusts knob settings and runs the workload. Although more samples improve tuning quality, they also bring significant time and computational costs. Thus, effective tuning with limited samples is a key challenge.

Clustering small, high-dimensional, sparse datasets is difficult. Traditional methods like K-Means rely on Euclidean distance, which is sensitive to noise and poorly captures local structures in sparse, high-dimensional data \cite{ahmed2020k}. Accurate modeling of such complex data is crucial for \systemname.

To address this, \systemname~employs a two-stage clustering strategy. For small sample sizes ($\leq \tau$), it uses cosine similarity–based spectral clustering \cite{von2007tutorial}, suited for sparsity and high dimensionality. For larger samples ($>\tau$), it applies Kernel PCA \cite{scholkopf1997kernel} followed by prototype-based clustering (e.g., K-medoids \cite{park2009simple}) to capture nonlinear structures and improve accuracy. In this paper, we set $\tau = 50$ according to empirical observations

This progressive approach enables \systemname~to robustly cluster across varying sample sizes, facilitating accurate structure discovery and enhancing tuning under sample constraints.

\textbf{Stage 1: Cosine Similarity and Spectral Clustering.}
When the sample size is below a threshold $\tau$, \systemname~employs a spectral clustering approach based on cosine similarity to address the challenges posed by sparse and high-dimensional data. Cosine similarity is defined as:
\begin{equation}
    \text{cosine\_sim}(x_i, x_j) = \frac{\langle x_i, x_j \rangle}{\|x_i\| \|x_j\|}, \quad where~ x_i = [\mathbf{r}_i, p(\mathbf{r}_i)],
\end{equation}
and is more effective than Euclidean distance in capturing directional similarity in sparse spaces.
To better model nonlinear structures, we construct a Gaussian-optimized similarity matrix:
\begin{equation}
    W_{ij} = \exp\left(-\gamma \cdot (1 - \text{cosine\_sim}(x_i, x_j))^2\right),
\end{equation}
where $\gamma = 1.0$ controls sensitivity to local variation and noise. Based on $W$, we compute the normalized Laplacian:
\begin{equation}
    L_{\text{norm}} = I - D^{-1/2} W D^{-1/2},
\end{equation}
where $D$ is the degree matrix. We then extract the two eigenvectors corresponding to the smallest eigenvalues of $L_{\text{norm}}$, forming a low-dimensional representation that preserves key structural relationships. Finally, K-Means is applied in this embedded space to produce the final clusters.

\textbf{Stage 2: Kernel PCA and K-Medoids Clustering.} 
When the sample size exceeds $\tau$, \systemname~applies Kernel PCA followed by K-Medoids. 
Each sample $x_i = [\mathbf{r}_i, p(\mathbf{r}_i)]$ is mapped into a high-dimensional space via the RBF kernel:
\begin{equation}
    K_{ij} = \exp\left(-\frac{\|x_i - x_j\|^2}{2\sigma^2}\right),
\end{equation}
where $\sigma$ is the bandwidth. After centering the kernel matrix, we perform eigen decomposition and retain the top $k$ principal components ($k \leq 5$) for dimensionality reduction. The resulting embeddings are clustered using K-Medoids, which iteratively selects medoids, assigns samples, and updates until convergence.

\subsubsection{Classification Method}
To partition the database knob tuning space, we adopt a Soft Margin SVM to address the challenges of noise and overlapping samples caused by the complex mapping between knob configurations and performance metrics. Unlike standard SVM, which seeks a hyperplane that perfectly separates classes, Soft Margin SVM introduces slack variables to allow margin violations, enabling robust classification in non-separable scenarios.
Formally, the optimization objective is:
\begin{equation}
\min_{w, b, \zeta} \ \frac{1}{2} \|w\|^2 + C \sum_{i=1}^{m} \zeta_i \quad \text{s.t.} \quad y_i (w^T x_i + b) \geq 1 - \zeta_i,\ \zeta_i \geq 0
\end{equation}
where $x_i = [\mathbf{r}_i, p(\mathbf{r}_i)]$ is the feature vector, $y_i$ is its label from the clustering step, and $C$ is a regularization parameter controlling the trade-off between margin size and misclassification penalty.

By allowing limited classification errors, Soft Margin SVM identifies a more flexible decision boundary, effectively partitioning the high-dimensional knob space even under noisy and partially overlapping data.

\subsection{Knob Tuning Workflow}
We present the complete tuning workflow of \systemname, which leverages a hierarchical and hybrid strategy to efficiently explore the vast configuration space. At its core lies a spatial decomposition algorithm that recursively partitions the high-dimensional space into manageable subregions. Within this structure, Monte Carlo Tree Search (MCTS) is employed to prioritize subregions with higher potential based on historical feedback, striking a balance between exploration and exploitation. Bayesian Optimization (BO) is then applied within selected subregions to perform fine-grained, model-guided sampling. This combination enables \systemname~to concentrate computational efforts on promising regions, reduce redundant evaluations, and significantly improve the chances of discovering optimal or near optimal configurations. The full procedure is summarized in Algorithm~\ref{ALG_2}.

\begin{algorithm}[t]
\caption{Knob Tuning via Spatial Decomposition and MCTS}
\label{ALG_2}
\renewcommand{\algorithmicrequire}{\textbf{Input:}}
\renewcommand{\algorithmicensure}{\textbf{Output:}}
\begin{algorithmic}[1]
\REQUIRE Configuration space $R$
\ENSURE Optimized knob setting $\mathbf{r}^\star$

\textcolor{gray}{\textit{// Cold Start Phase}}
\STATE Initialize by sampling $n$ configurations $\{\mathbf{r}_j\}_{j=1}^n \subset R$
\STATE Initialize the sample set $D_0 \leftarrow \emptyset$
\FOR{each iteration $i = 1$ to $n$}
    \STATE Evaluate $\mathbf{r}_i$ to obtain its performance $p(\mathbf{r}_i)$
    \STATE Update $D_i \leftarrow D_{i-1} \cup \{(\mathbf{r}_i, p(\mathbf{r}_i))\}$
\ENDFOR

\textcolor{gray}{\textit{// Tuning Loop}}
\FOR{iteration $i = n+1, 2, \dots, N$}
    \STATE Initialize node $A \leftarrow$ root, region $R_A \leftarrow R$
    \WHILE{$A$ is splittable}
        \STATE $(R_B, R_C) \leftarrow$ \textsc{Spatial\_Decomposition}$(D_{i-1}, R_A)$
        \STATE Compute $\text{UCB}_B$, $\text{UCB}_C$ 
        \IF{$\text{UCB}_B < \text{UCB}_C$}
            \STATE Set node $A \leftarrow$ node $C$, $R_A \leftarrow R_C$.
        \ELSE
            \STATE Set node $A \leftarrow$ node $B$, $R_A \leftarrow R_B$.
        \ENDIF
    \ENDWHILE
    \STATE $R_{\text{selected}} \leftarrow R_A$
    \STATE Use BO to propose a new configuration $\mathbf{r}_i$ within $R_{\text{selected}}$
    \STATE Evaluate $\mathbf{r}_i$ to obtain its performance $p(\mathbf{r}_i)$
    \STATE Update $D_i \leftarrow D_{i-1} \cup \{(\mathbf{r}_i, p(\mathbf{r}_i))\}$
\ENDFOR
\STATE \textbf{return} $\mathbf{r}^\star = \arg\max_{\mathbf{r} \in D_N} p(\mathbf{r})$
\end{algorithmic}
\end{algorithm}

At the start of the tuning process, \systemname~enters a cold-start phase during the first $n$ iterations to collect initial performance observations. Effectively handling this stage is crucial, as the initial samples significantly influence the subsequent optimization trajectory.

Following prior work~\cite{kanellis2022llamatune,zhang2021facilitating,lao2023GPTuner}, we generate $n$ initial configurations $\{\mathbf{r}_j\}_{j=1}^n \subset R$ using Latin Hypercube Sampling (LHS)~\cite{mckay1992latin}, which ensures uniform coverage of the configuration space. These configurations are evaluated to obtain performance scores $\{p(\mathbf{r}_j)\}_{j=1}^n$, forming the initial sample set $D_n = \{(\mathbf{r}_j, p(\mathbf{r}_j))\}_{j=1}^n$.

After initialization, \systemname~enters its iterative tuning phase based on spatial decomposition. In each iteration, the search tree is reconstructed from the root, and the configuration space is recursively partitioned. For each node $A$, the algorithm maintains its region $R_A$ and associated statistics, including sample count $n_A$ and average performance $v_A$ (see Fig.\ref{FIG_4}).

To avoid over-exploiting high-performing but potentially suboptimal regions, \systemname~uses the Upper Confidence Bound (UCB) strategy to guide node selection. UCB balances exploration and exploitation by considering both performance and visitation frequency. The UCB score for a child node $B$ is defined as:
\begin{equation}
\text{UCB}_B = \frac{v_B}{n_B} + 2 C_p \cdot \sqrt{\frac{2 \log n_A}{n_B}}
\end{equation}
where $C_p$ is a tunable exploration parameter, $n_A$ is the parent node’s visit count, and $v_B$, $n_B$ denote the cumulative value and visit count of node $B$, respectively. The child with the highest UCB score is selected for further expansion. This strategy enables \systemname~to concentrate on promising regions while preserving global awareness, effectively mitigating premature convergence.

Once an exploration region ($R_{\text{selected}}$) is selected via UCB-guided traversal, the path from the root to the target leaf defines a constrained subregion as the intersection of soft-margin SVM decision boundaries. Within this region, \systemname~employs BO to perform localized sampling and refine the search for high-performing configurations.

In this context, BO plays a supporting role by enabling fine-grained optimization within each selected partition, effectively complementing the global partitioning strategy. By restricting BO to promising subspaces, \systemname~avoids ineffective global exploration and accelerates convergence toward the optimum.

In summary, \systemname~addresses the challenges of high-dimensional and nonlinear knob tuning by combining recursive spatial decomposition with UCB-driven adaptive search. The integration of constrained BO sampling further enhances efficiency by focusing optimization efforts within the most promising regions.

\section{Experiment}
\begin{figure*}[htbp]
	\centering
	\begin{minipage}[c]{0.98\textwidth}
		\centering
		\includegraphics[width=\textwidth]{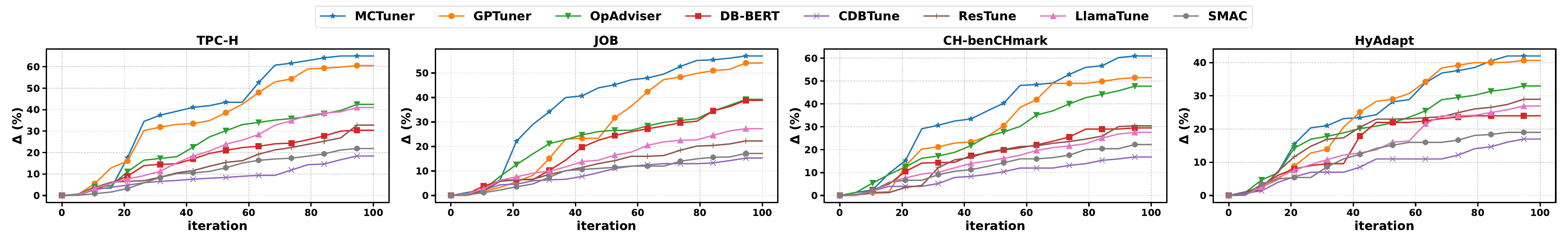}
        \vspace{-1.7em}
		\subcaption{Latency analysis.}
		\label{fig_E2_1}
	\end{minipage} \\
	\begin{minipage}[c]{0.74\textwidth}
		\centering
		\includegraphics[width=\textwidth]{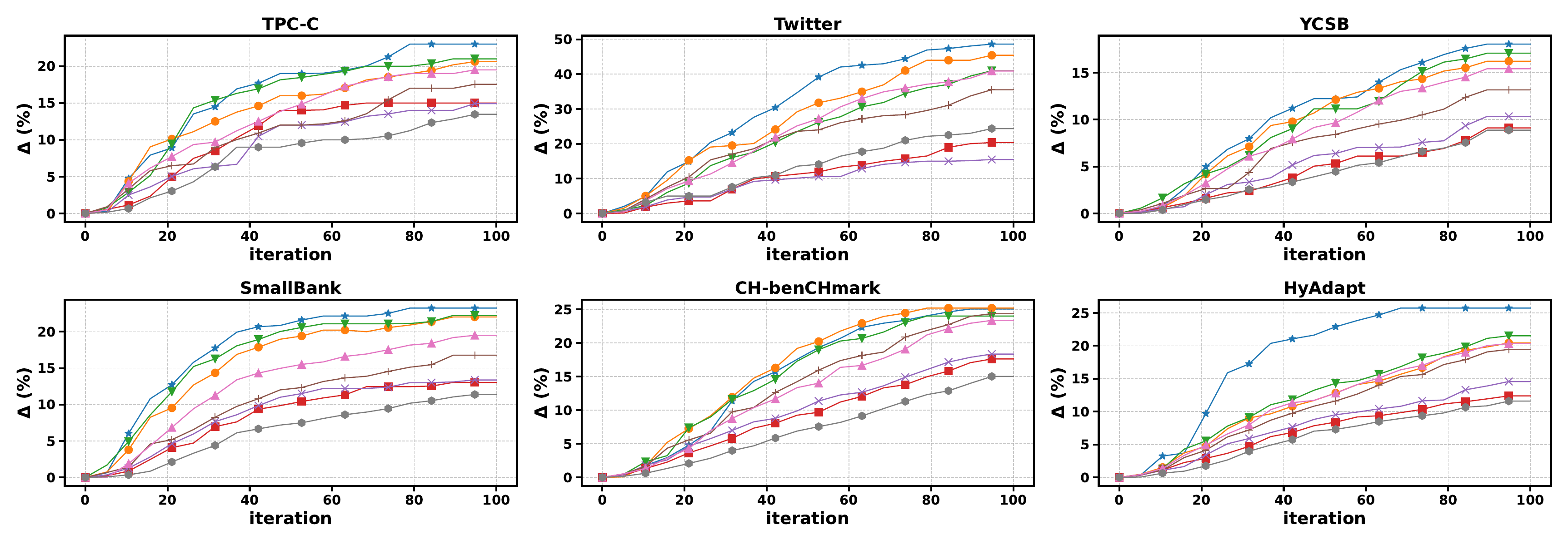}
        \vspace{-1.7em}
		\subcaption{Throughput analysis.}
		\label{fig_E2_2}
	\end{minipage} 
    \begin{minipage}[c]{0.24\textwidth}
		\centering
		\includegraphics[width=\textwidth]{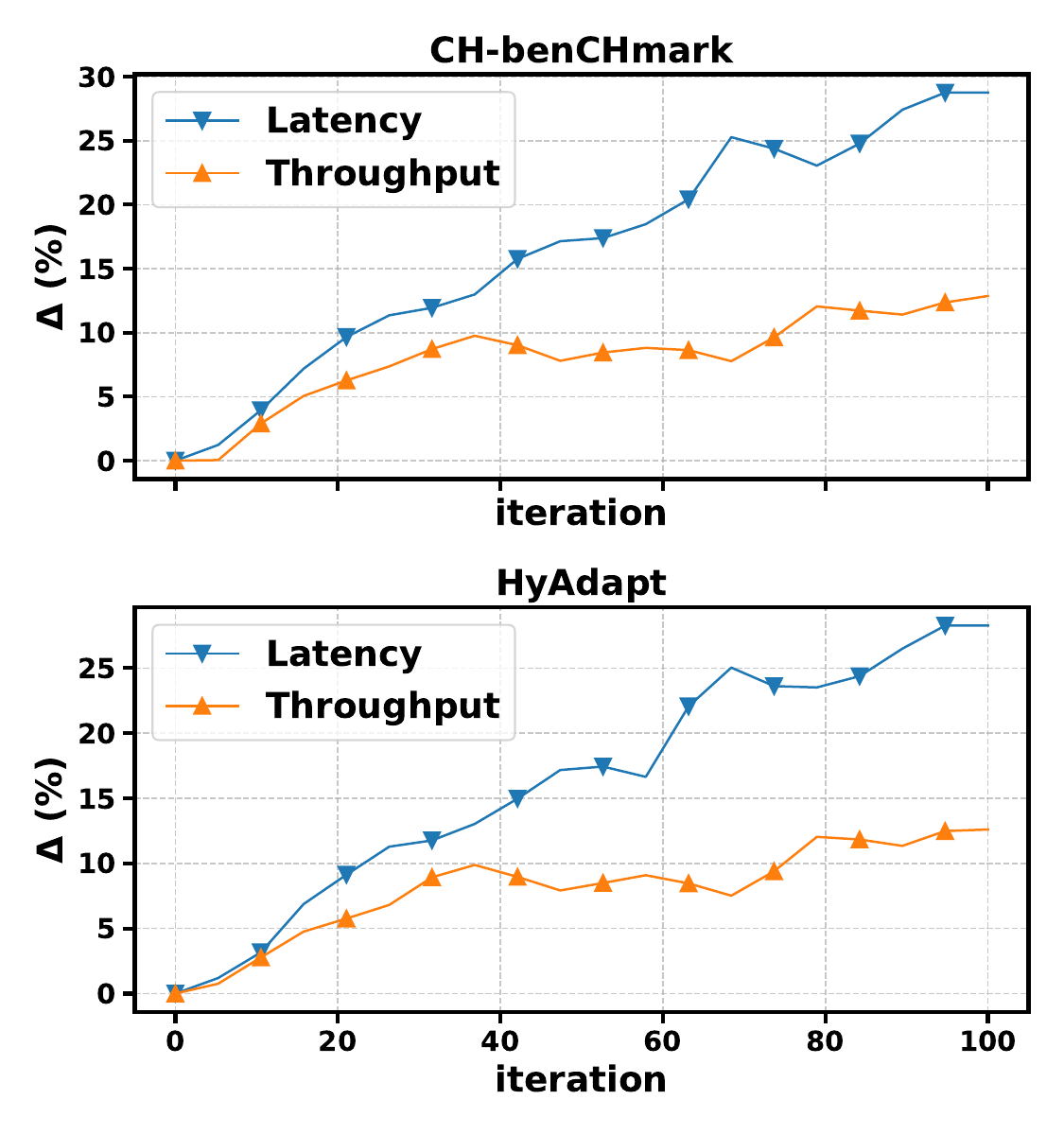}
        \vspace{-1.7em}
		\subcaption{Multi-purpose analysis.}
		\label{fig_E2_3}
	\end{minipage} 
         \vspace{-1em}
	\caption{Performance improvement over iterations across 8 benchmarks (top-right is better).}
	\label{fig_E2}
\end{figure*}
\subsection{Experiment Settings}
\subsubsection{Target Workload}
We selected eight benchmarks for our experiment: two OLAP workloads (TPC-H, JOB), four OLTP workloads (TPC-C, Twitter, YCSB, SmallBank), and two HTAP workloads (CH-benCHmark, HyAdapt). These benchmarks represent a broad spectrum of database operations. All implementations were sourced from benchmark \cite{DifallahPCC13,leis2015good}, known for its accurate standard benchmark implementations, ensuring the consistency and validity of our setup.
\subsubsection{Hardware}
Our experimental procedures are executed on a cloud server furnished with a 24-core Intel Xeon E5-2676 v3 CPU, 128 GB of Random Access Memory (RAM), and a 1TB Solid State Drive (SSD). Moreover, a GeForce RTX 2080 Ti graphics card with 22 GB of dedicated memory is also utilized in our experiments. 
\subsubsection{Baseline}
\systemname~is implemented in Python using the GPT-4-powered OpenAI API. We then conducted a comparative analysis against the following state-of-the-art methods:

\begin{itemize}[leftmargin=*]
    \item \textbf{GPTuner} \cite{lao2023GPTuner} employs GPT-guided knowledge summarization and coarse-to-fine BO to optimize the configurable space.
    \item \textbf{ResTune} \cite{zhang2021restune} converts resource-oriented configuration tuning into a constrained BO problem, using historical task data and meta-learning to accelerate cloud database tuning.
    \item \textbf{LlamaTune} \cite{kanellis2022llamatune} improves sample efficiency by narrowing the configurable space through random projections, biased sampling, and knob bucketing, thereby optimizing configurations across database versions.
    \item \textbf{SMAC} \cite{hutter2011sequential} is a BO method using random forest as the surrogate model, excelling in knob tuning.
    \item \textbf{DB-BERT} \cite{trummer2022db} uses BERT for tuning hints extraction and reinforcement learning with feedback to optimize database knobs.
    \item \textbf{CDBTune} \cite{zhang2019end} is an end-to-end cloud database tuning system using deep reinforcement learning and trial-and-error strategies to find optimal configurations.
    \item \textbf{OpAdviser} \cite{zhang2023efficient} automatically creates a compact configurable space and selects an optimizer based on historical data, speeding up database tuning and reducing runs.
\end{itemize}

In conclusion, GPTuner, ResTune, LlamaTune, and SMAC are BO-based methods, while DB-BERT and CDBTune are RL-based methods. Additionally, OpAdviser plays a significant role in facilitating the tuning process by automatically selecting the most appropriate tuning method.

\subsubsection{Metrics}
We adopt relative performance improvement as the primary evaluation metric \cite{huang2025e2etuneendtoendknobtuning}. For OLAP workloads, the objective is to minimize query latency, and the improvement is calculated as: $\Delta = \frac{lat_{\text{default}}-lat}{lat_{\text{default}}}$, where $lat$ and $lat_{\text{default}}$ denote the latency under the optimized and default configurations, respectively.
For OLTP workloads, the objective is to maximize throughput (transactions per second, $tps$), and the improvement is defined as:
$\Delta = \frac{tps - tps_{\text{default}}}{tps_{\text{default}}}$, where $tps$ and $tps_{\text{default}}$ represent the throughput under the optimized and default configurations, respectively. For HTAP workloads, both latency and throughput improvements are reported.

\subsubsection{Tuning Setting}

We conducted experiments using PostgreSQL v14.9 and gpt-4o-mini. Based on prior research \cite{kanellis2022llamatune, zhang2019end, lao2023GPTuner}, we initially configured 60 knobs identically across all algorithms, and then fine-tuned them according to each optimization method. For example, DB-BERT and GPTuner leverage LLMs for knowledge extraction, while OpAdviser refines the configurable space based on insights from similar historical workloads. To ensure a fair comparison under controlled computational budgets, each method was limited to 100 iterations, with one configuration sampled and evaluated per iteration. We optimized throughput for OLTP and HTAP workloads, and the 95th percentile latency for OLAP and HTAP workloads, using performance improvement as the key metric. Additionally, after each iteration, the target workload was executed, and once a method proposed a knob setting, the database was restarted, as some configurations only take effect after a restart.

For BO-based methods (e.g., \systemname, SMAC), the first 10 iterations were allocated for cold-start sampling using LHS. The remaining 90 iterations focused on optimizing the configuration. Furthermore, LlamaTune was integrated with SMAC as the recommended optimizer, and the CPU metric in ResTune was replaced with average latency.
For RL-based methods, we followed recent work (e.g., DB-BERT, CDBTune) and avoided training neural networks, as evaluations indicated that trained networks may suffer from overfitting \cite{zhang2021facilitating}. The model for selecting tuning methods in OpAdviser was trained using our own collected dataset. In the main results, \systemname~uses a GP-based BO surrogate model. 
\vspace{-0.2em}
\subsection{Performance Comparison}

Fig.~\ref{fig_E2} shows that \systemname~outperforms other methods in both latency and throughput as iterations increase, quickly identifying high-quality knob settings and adapting well to diverse workloads.


\subsubsection{Latency Analysis}
As shown in Fig.~\ref{fig_E2_1}, \systemname~consistently outperforms other methods, maintaining the lowest latency across most iterations. Moreover, it quickly adapts to different workloads, identifying a high-performance knob setting within the first 50 iterations (50 samples), demonstrating strong adaptability with a small number of samples.
For most workloads, the latency reduction ($\Delta$) steadily decreases, reaching an optimal value around the 80th iteration, which highlights \systemname's ability to quickly optimize system performance. In contrast, GPTuner achieves a similar reduction in latency during the first 40 iterations but lags slightly behind due to its less efficient knob search strategy. While GPTuner continues to reduce latency, it remains higher than \systemname~overall.
OpAdviser performs well in the early stages, with latency reduction comparable to \systemname~ and GPTuner, particularly on the CH-benCHmark and HyAdapt workloads. However, its performance diminishes in subsequent iterations, likely attributable to its automated optimizer selection strategy, which lacks the necessary adaptability during the later phases of the tuning process.

Other methods, such as DB-BERT, ResTune, and LlamaTune, show strong initial performance but plateau in later stages, resulting in diminishing returns. CDBTune struggles in early iterations, with minimal latency reduction, suggesting that its DDPG-based configuration method faces challenges adapting to different workloads. Although latency improves in later iterations, it remains higher than \systemname’s, reflecting its limited optimization capacity. SMAC follows a similar trend, with slow latency reduction in early iterations, likely due to its weaker exploration ability in high-dimensional knob spaces. While latency improves in later iterations, it still remains significantly higher than \systemname’s, indicating lower optimization efficiency.

In summary, \systemname~consistently outperforms the strongest competing method across all evaluated workloads, achieving relative performance gains of 6.6\% on TPC-H, 5.4\% on JOB, 19.2\% on CH-benCHmark, and 2.4\% on HyAdapt.

\subsubsection{Throughput Analysis}
In the throughput analysis (Fig.~\ref{fig_E2_2}), \systemname~achieves the highest throughput across most workloads, particularly TPC-C, SmallBank, and HyAdapt, further demonstrating its adaptability to diverse workloads with minimal samples.
Its throughput growth follows a nearly linear trend in the first 50 iterations, reaching its optimal value around the 80th iteration, indicating efficient and sustained optimization. This performance is likely due to \systemname’s optimization strategy, which maintains high processing capacity while handling more requests.

OpAdviser and GPTuner also show strong throughput improvement, with OpAdviser excelling on SmallBank and TPC-C, though its throughput is slightly lower than \systemname's. GPTuner performs well on workloads with higher processing demands, such as Twitter. LlamaTune ranks just below these methods, particularly excelling on TPC-C and SmallBank due to its space compression algorithm, but its final performance falls short of \systemname's. ResTune shows notable throughput improvement on workloads like CH-benCHmark and TPC-C, benefiting from its Bayesian Optimization approach. While DB-BERT excels in latency, it shows minimal throughput improvement, lagging behind other methods. SMAC and CDBTune face similar challenges, showing some throughput improvement but underperforming compared to \systemname~and other more efficient methods.

In summary, \systemname~demonstrates consistently strong performance across diverse workloads, achieving improvements of 9.4\% on TPC-C, 6.2\% on Twitter, 12.3\% on YCSB, 4.6\% on SmallBank, and 13.6\% on HyAdapt.

\subsubsection{Multi-Purpose Analysis}
In our experiments, we conducted multi-objective optimization for HTAP workloads, specifically CH-benCHmark and HyAdapt, with MoE assigning a weight of $0.6$ to latency and $0.4$ to throughput. The weights are assigned based on the need to prioritize low latency for real-time tasks while also optimizing throughput for batch tasks in hybrid workload scenarios. The corresponding experimental results are shown in Fig.~\ref{fig_E2_3}. The results demonstrate notable enhancements in both latency and throughput, although these improvements are marginally lower than those achieved by single-objective optimization. Furthermore, some iterations exhibit a trade-off, where latency improvement is accompanied by a decline in throughput, which arises from the multi-objective optimization approach. Despite these interactions, our method achieves overall optimization improvements.

\subsubsection{Tuning Efficiency}
\begin{table}
\caption{Algorithm time per iteration on average.}
\begin{center}
\setlength{\tabcolsep}{4pt}
\begin{tabular}{cccc}
\rowcolor{gray!20}
\toprule
\systemname&GPTuner&OpAdviser&ResTune\\
\midrule
5.54s & 7.47s & 6.39s & 8.22s \\
\rowcolor{gray!20}
\midrule
LlamaTune &SMAC & DB-BERT& CDBTune\\
\midrule
2.06s & 1.87s & 14.61s & 0.54s \\
\bottomrule
\end{tabular}
\label{time}
\end{center}
\vspace{-1em}
\end{table}

\begin{table*}[t]
\centering
\footnotesize
\renewcommand{\arraystretch}{1.2}
\setlength{\tabcolsep}{3pt}
\caption{Evaluation of configuration space compression.
\textit{ratio$_1$} indicates the compression ratio achieved by the MoE module relative to the original configuration space; \textit{ratio$_2$} denotes the size of the final sampling space, after applying the spatial decomposition algorithm, as a proportion of the MoE-compressed space. Knobs selected are color-coded according to their respective methods. The performance improvement obtained using BO are reported in the last row.}
\begin{adjustbox}{max width=\textwidth}
\begin{tabular}{cc|ccc|cc|ccc}
\toprule
\multicolumn{5}{c|}{\textbf{OLTP (TPC-C)}} & 
\multicolumn{5}{c}{\textbf{OLAP (TPC-H)}} \\
\cmidrule(lr){1-5} \cmidrule(lr){6-10} 
\textbf{SHAP} & \textbf{\textcolor{ForestGreen}{CART}} & \textbf{\textcolor{red}{\textbf{MoE}}} & \textbf{\textit{ratio$_1$}} & \textbf{\textit{ratio$_2$}} & \textbf{SHAP} & \textbf{\textcolor{ForestGreen}{CART}} & \textbf{\textcolor{red}{\textbf{MoE}}} & \textbf{\textit{ratio$_1$}} & \textbf{\textit{ratio$_2$}} \\
\midrule
random\_page\_cost & \textcolor{ForestGreen}{commit\_siblings} & \textcolor{red}{effective\_io\_concurrency} & $1.0\%$ & $0.4\%$ &
random\_page\_cost & maintenance\_work\_mem & effective\_io\_concurrency  & $19.9\%$ & $2.8\%$ \\
join\_collapse\_limit & \textcolor{ForestGreen}{commit\_delay} & \textcolor{red}{random\_page\_cost} & $<0.1\%$ & $23.4\%$ &
seq\_page\_cost & \textcolor{ForestGreen}{seq\_page\_cost} & shared\_buffers & $<0.1\%$ & $<0.1\%$ \\
checkpoint\_completion\_target & \textcolor{ForestGreen}{checkpoint\_completion\_target} & shared\_buffers & $<0.1\%$ & $<0.1\%$ &
checkpoint\_completion\_target & \textcolor{ForestGreen}{random\_page\_cost} & \textcolor{red}{random\_page\_cost} & $<0.1\%$ & $86.5\%$ \\
commit\_siblings & \textcolor{ForestGreen}{max\_replication\_slots} & \textcolor{red}{autovacuum\_vacuum\_scale\_factor} & $<0.1\%$ & $93.4\%$ &
commit\_siblings & wal\_buffers & wal\_writer\_flush\_after & $0.2\%$ & $1.8\%$ \\
autovacuum\_analyze\_scale\_factor & checkpoint\_timeout & \textcolor{red}{max\_parallel\_workers\_per\_gather} & $2.2\%$ & $5.1\%$ &
autovacuum\_analyze\_scale\_factor & commit\_delay & \textcolor{red}{autovacuum\_max\_workers} & $<0.1\%$ & $18.3\%$ \\
seq\_page\_cost & min\_wal\_size & work\_mem & $<0.1\%$ & $<0.1\%$ &
autovacuum\_vacuum\_scale\_factor & \textcolor{ForestGreen}{autovacuum\_max\_workers} & \textcolor{red}{autovacuum\_vacuum\_scale\_factor} & $0.9\%$ & $99.9\%$ \\
bgwriter\_lru\_multiplier & wal\_writer\_delay & checkpoint\_flush\_after & $43.8\%$ & $0.3\%$ &
from\_collapse\_limit & \textcolor{ForestGreen}{max\_parallel\_workers\_per\_gather} & checkpoint\_flush\_after & $12.5\%$ & $1.9\%$ \\
autovacuum\_vacuum\_scale\_factor & \textcolor{ForestGreen}{max\_parallel\_workers\_per\_gather} & \textcolor{red}{wal\_buffers} & $6.2\%$ & $<0.1\%$ &
join\_collapse\_limit & \textcolor{ForestGreen}{min\_wal\_size} & wal\_buffers & $<0.1\%$ & $<0.1\%$ \\
max\_parallel\_workers\_per\_gather & \textcolor{ForestGreen}{join\_collapse\_limit} & \textcolor{red}{commit\_siblings} & $100.0\%$ & $6.3\%$ &
autovacuum\_max\_workers & work\_mem & \textcolor{red}{backend\_flush\_after} & $50.0\%$ & $1.4\%$ \\
from\_collapse\_limit & \textcolor{ForestGreen}{autovacuum\_analyze\_scale\_factor} & \textcolor{red}{seq\_page\_cost} & $<0.1\%$ & $28.2\%$ &
vacuum\_cost\_limit & effective\_cache\_size & \textcolor{red}{seq\_page\_cost} & $<0.1\%$ & $97.6\%$ \\
commit\_delay & wal\_writer\_flush\_after & \textcolor{red}{commit\_delay} & $5.0\%$ & $<0.1\%$ &
max\_parallel\_workers\_per\_gather & \textcolor{ForestGreen}{max\_wal\_senders} & commit\_delay & $4.0\%$ & $<0.1\%$ \\
vacuum\_cost\_limit & \textcolor{ForestGreen}{random\_page\_cost} & \textcolor{red}{bgwriter\_lru\_multiplier} & $<0.1\%$ & $8.2\%$ &
bgwriter\_lru\_multiplier & wal\_writer\_delay & \textcolor{red}{bgwriter\_lru\_multiplier} & $80.0\%$ & $42.0\%$ \\
vacuum\_cost\_delay & \textcolor{ForestGreen}{log\_temp\_files} & checkpoint\_timeout & $4.1\%$ & $<0.1\%$ &
max\_wal\_senders & \textcolor{ForestGreen}{from\_collapse\_limit} & \textcolor{red}{max\_parallel\_workers\_per\_gather} & $2.2\%$ & $19.8\%$ \\
autovacuum\_max\_workers & \textcolor{ForestGreen}{autovacuum\_max\_workers} & \textcolor{red}{checkpoint\_completion\_target} & $100.0\%$ & $99.9\%$ &
min\_wal\_size & default\_statistics\_target & \textcolor{red}{max\_parallel\_workers} & $2.2\%$ & $7.2\%$ \\
max\_replication\_slots & \textcolor{ForestGreen}{from\_collapse\_limit} & \textcolor{red}{from\_collapse\_limit} & $<0.1\%$ & $5.5\%$ &
backend\_flush\_after & \textcolor{ForestGreen}{autovacuum\_vacuum\_scale\_factor} & \textcolor{red}{checkpoint\_completion\_target} & $100.0\%$ & $99.9\%$ \\
bgwriter\_lru\_maxpages & \textcolor{ForestGreen}{wal\_buffers} & effective\_cache\_size & $<0.1\%$ & $<0.1\%$ &
bgwriter\_lru\_maxpages & max\_connections & \textcolor{red}{min\_wal\_size}& $<0.1\%$ & $<0.1\%$ \\
max\_worker\_processes & \textcolor{ForestGreen}{seq\_page\_cost}& maintenance\_work\_mem & $<0.1\%$ & $<0.1\%$ &
max\_parallel\_workers & effective\_io\_concurrency & \textcolor{red}{bgwriter\_lru\_maxpages} & $<0.1\%$ & $<0.1\%$ \\
log\_temp\_files & max\_wal\_senders & \textcolor{red}{max\_worker\_processes} & $<0.1\%$ & $1.3\%$ &
max\_replication\_slots & \textcolor{ForestGreen}{vacuum\_cost\_limit} & \textcolor{red}{join\_collapse\_limit} & $<0.1\%$ & $27.2\%$ \\
wal\_buffers & max\_parallel\_workers & \textcolor{red}{join\_collapse\_limit} & $<0.1\%$ & $5.5\%$ &
max\_standby\_streaming\_delay & \textcolor{ForestGreen}{max\_parallel\_workers} & \textcolor{red}{from\_collapse\_limit} & $<0.1\%$ & $34.1\%$ \\
effective\_io\_concurrency & effective\_cache\_size & \textcolor{red}{bgwriter\_lru\_maxpages} & $80.0\%$ & $<0.1\%$ &
log\_rotation\_size & checkpoint\_flush\_after & effective\_cache\_size & $6.2\%$ & $<0.1\%$ \\
\rowcolor{gray!20}
\midrule
$8.7\%$ & $7.8\%$ & $\mathbf{12.9\%}\uparrow$ & - & - &
$24.3\%$ & $10.0\%$ & $\mathbf{25.2\%}\uparrow$ & - & - \\

\bottomrule
\end{tabular}
\end{adjustbox}
\label{knob_selection}
\end{table*}

The data presented in Table~\ref{time} reflects only the time required by the algorithm to determine a new set of evaluation knobs, excluding workload execution time. \systemname~demonstrates exceptional tuning efficiency, requiring just 5.54 seconds, significantly outperforming most other methods. While GPTuner and OpAdviser achieve similar performance to \systemname, they are slightly less efficient, suggesting they take more time to reach comparable results. Although LlamaTune, SMAC, and CDBTune demonstrate high algorithmic efficiency, they fall significantly short of \systemname~in performance improvement, showing more than a 50\% deficit in overall gains. LlamaTune and SMAC also utilize BO, but \systemname~enhances this with a space decomposition strategy, resulting in slightly longer tuning times but superior overall performance. In conclusion, \systemname~strikes an optimal balance between tuning efficiency and performance improvement, demonstrating robust optimization capabilities.


\vspace{-1em}
\subsection{Space Compression Insights}

\subsubsection{Effectiveness of Knob Selection}
We analyze the knob selection strategy employed by MCTuner. We conduct experiments on both OLTP and OLAP workloads using TPC-C and TPC-H benchmarks, respectively. We compare MCTuner's mixture-of-experts (MoE) approach against several baselines, including CART, LASSO, and a zero-shot LLM-based method. CART and LASSO are provided with knob-setting data from 50 iterations to ensure a fair comparison. The final evaluation is performed using SHAP values to assess the quality of the selected knobs.
The results based on the top 20 selected knobs are presented in Table~\ref{knob_selection}, while detailed analyses for LASSO and the zero-shot LLM are deferred to the Appendix. Knob names highlighted in the same color as a method denote overlap with those identified by SHAP.

Across both TPC-C and TPC-H workloads, MoE consistently demonstrates superior performance in balancing effectiveness, efficiency, and interpretability. On TPC-C, MoE identifies 14 SHAP-overlapping knobs, outperforming CART (13) and LASSO (8), while completing selection in only 389.5 seconds, compared to over 3400 seconds for CART and LASSO. On TPC-H, MoE similarly achieves 13 overlaps, surpassing CART and LASSO (both with 10), and requires just 373.3 seconds, significantly faster than the 650$+$ seconds needed by the baselines. Although the zero-shot LLM method finishes in under 10 seconds on both workloads, it achieves only 11 and 10 overlapping knobs on TPC-C and TPC-H respectively, and falls short in accuracy. In addition, MoE provides interpretable rationales for selected knobs, enhancing transparency and making it more suitable for practical deployment.

We evaluate the selected knobs in a downstream tuning task using Bayesian Optimization with Gaussian Processes, with all knob ranges standardized based on the zero-shot LLM for fairness (results in Table~\ref{knob_selection}). MoE consistently outperforms other methods across both TPC-C and TPC-H workloads, achieving 12.9\% and 25.2\% improvements, respectively. While LASSO yields slightly higher improvement on TPC-C (14.3\%), it incurs significantly higher cost (3404.9s) and lacks interpretability. In contrast, MoE balances performance and efficiency while providing interpretable rationales. It also identifies impactful knobs overlooked by SHAP, demonstrating its ability to capture non-obvious yet effective configurations. These results confirm MoE’s practical advantage in tuning accuracy, runtime efficiency, and transparency, making it well-suited for real-world deployment.

Overall, both the zero-shot LLM and MoE-based approaches consistently identify high-impact knobs across diverse workloads. Crucially, as these methods do not depend on workload-specific performance feedback during selection, they significantly reduce the cost of knob selection, facilitating efficient and effective tuning.

\subsubsection{Space Compression Ratio}

We evaluate the effectiveness of \systemname~in compressing the configuration space during its two-stage optimization process, as summarized in Table~\ref{knob_selection}.



As shown in Table~\ref{knob_selection}, \systemname~achieves substantial compression of the configuration space, highlighting its effectiveness in narrowing the tuning domain. Notably, certain knobs such as \texttt{checkpoint\_completion\_target} show limited compression across both TPC-C and TPC-H, likely due to their already compact and well-defined original range $[0,1]$. Similar behavior is observed for \texttt{autovacuum\_vacuum\_scale\_factor} in both workloads, and for \texttt{seq\_page\_cost} and \texttt{random\_page\_cost} in TPC-H, where the MoE module already defines a narrow, high-quality search space.

These results demonstrate that \systemname~can efficiently compress the configuration space without sacrificing critical tuning potential, enabling scalable and targeted optimization.

\vspace{-0.2em}
\subsection{Ablation Study}
\subsubsection{Adaptability in Dynamic Workloads}
To evaluate the adaptability and robustness of \systemname~under dynamic workload conditions, we design a cyclic workload drift experiment based on TPC-C. The workload scale periodically varies following the sequence: Scale Factor (SF): $0.1\rightarrow1\rightarrow10\rightarrow1\rightarrow0.1$, simulating realistic fluctuations such as promotional surges or diurnal access patterns. Each phase is allocated 40 tuning iterations. In the initial phase (SF = 0.1), the first 10 iterations are initialized using LHS to address the cold-start problem. In subsequent phases, we adopt a warm-start strategy by seeding the search with the top-10 configurations obtained from the preceding phase.

\begin{figure}
\centering
\includegraphics[width=1\linewidth]{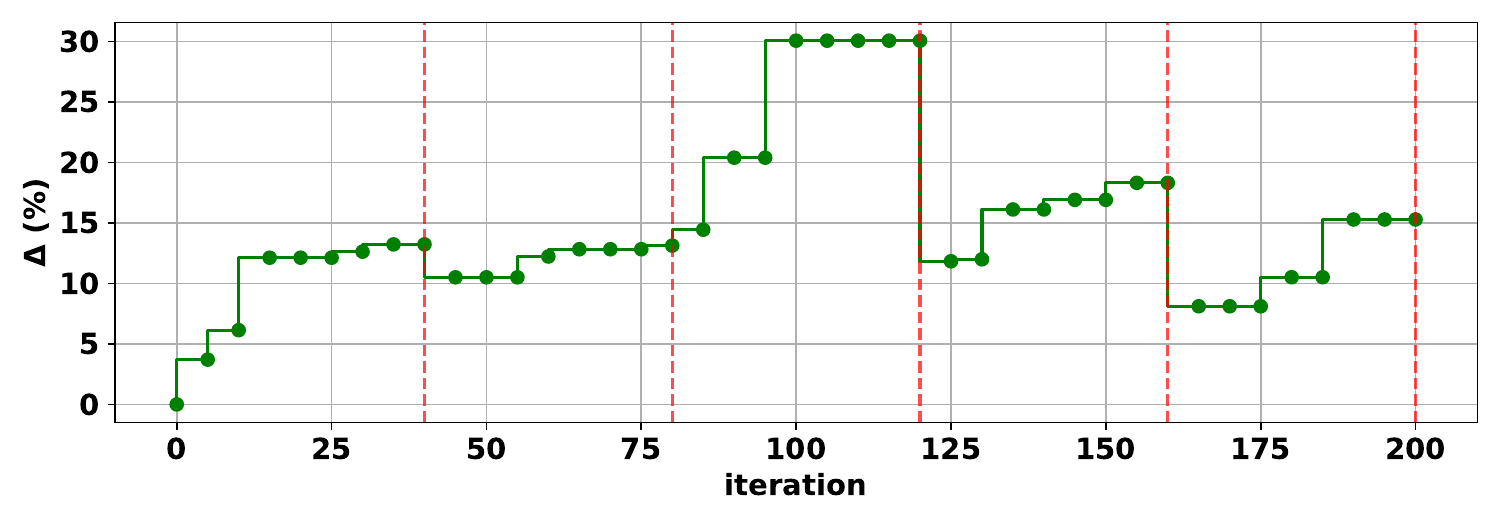}
\vspace{-2em}
\caption{Ablation study on workload drift.}
\vspace{-1em}
\label{FIG_wd}
\end{figure}

Fig.~\ref{FIG_wd} depicts the tuning trajectory across all workload phases. Green markers indicate the best configuration observed at each iteration, while red dashed lines denote workload transitions. Table~\ref{TAB_warmstart} summarizes the quantitative benefits of configuration transfer across phases, demonstrating that \systemname~effectively leverages historical knowledge to accelerate tuning under changing workload conditions.

\renewcommand{\arraystretch}{0.9}  
\begin{table}
\centering
\caption{Warm-start effectiveness across workload scale.}
\label{TAB_warmstart}
\begin{tabular}{cccc}
\toprule
\textbf{Transition (SF)} & \textbf{Gain (\%)} & \textbf{\#Configs $>$5\%} & \textbf{\#Configs $>$10\%} \\
\midrule
0.1 $\rightarrow$ 1   & 10.5 & 6 & 1 \\
1 $\rightarrow$ 10    & 20.4 & 8 & 5 \\
10 $\rightarrow$ 1    & 12.0 & 5 & 5 \\
1 $\rightarrow$ 0.1   & 8.1  & 2 & 0 \\
\bottomrule
\end{tabular}
\vspace{-1em}  
\end{table}

Notably, \systemname~consistently recovers performance shortly after each workload shift. For instance, when transitioning from SF 0.1 to 1, warm-starting with prior configurations results in an immediate improvement of up to 10.5 percent, which is further enhanced through continued tuning. An even more significant performance boost of 20.4 percent is observed when the scale factor changes from 1 to 10. This substantial gain not only confirms the transferability of the selected knobs by \systemname, but also highlights the influence of differing workload characteristics on tuning effectiveness. Specifically, the default throughput at SF equals 1 is 600.082, while at SF equals 10 it reaches 2683.464, which is more than four times higher, leading to distinct tuning dynamics. Furthermore, the tuning process across the two phases with SF equals 0.1 demonstrates sustained cross-phase optimization, a pattern similarly evident between the two SF equals 1 phases.

Interestingly, the configurations yielding the greatest improvements after workload transitions are not always those with the best prior phase performance. This indicates that while configuration transferability exists, it is not uniformly consistent. Therefore, continued online tuning remains crucial to fully adapt to evolving workload regimes.

\subsubsection{Ablations on BO}
\begin{figure}
\centering
\includegraphics[width=1\linewidth]{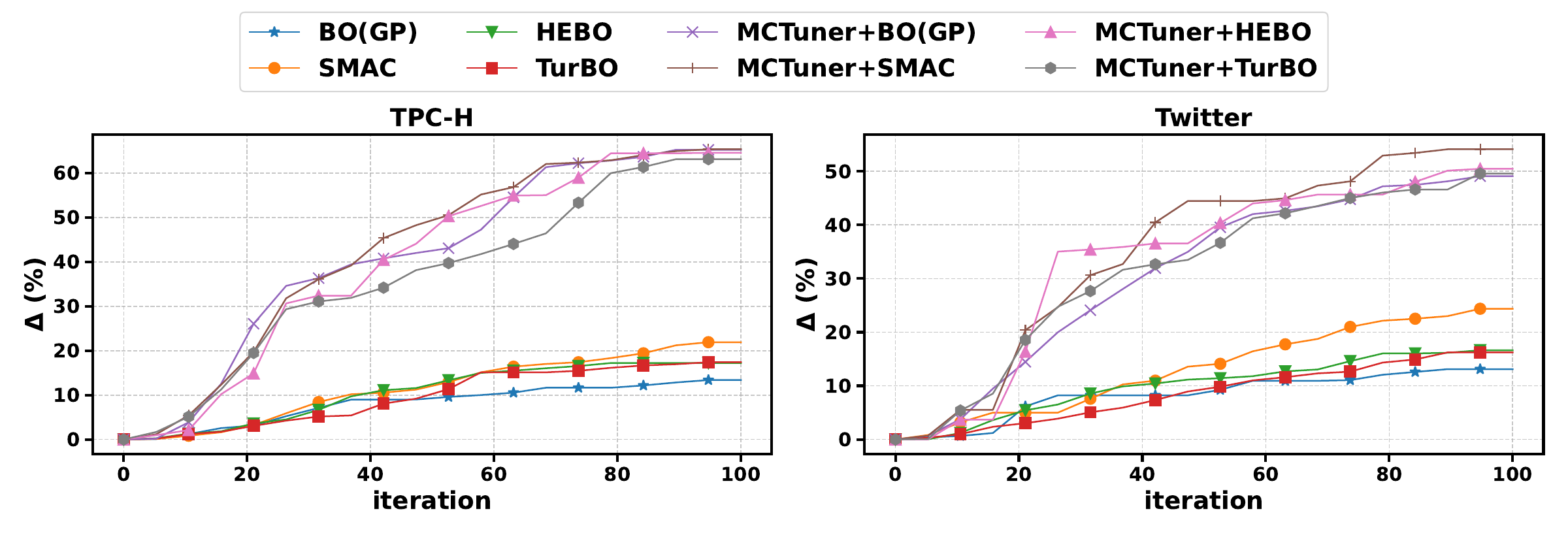}
\vspace{-2em}
\caption{Ablation study on BO.}
\vspace{-1em}
\label{FIG_BO}
\end{figure}

In the spatial decomposition algorithm, leaf nodes use BO to sample the next evaluation point. In our experiments, we employed the basic BO algorithm with a GP surrogate model. We then compared three alternative methods: SMAC, TurBO, and HEBO. SMAC excels in database knob tuning and does not require a continuous knob space, while TurBO is effective in hyperknob tuning, particularly in areas like Neural Architecture Search. HEBO is primarily used in E2ETune for training data collection. The experimental results are shown in Fig.~\ref{FIG_BO}.

As shown in the figure, SMAC outperforms the other original BO algorithms, as expected, while GP performs the worst due to its simpler assumptions. HEBO and TurBO show no significant differences. However, after incorporating \systemname, tuning performance improved substantially, further validating the approach proposed in the introduction that `a large portion of the configurable space is effectively meaningless'. Notably, the spatial decomposition algorithm minimized differences among the four methods, leading to similar final performances. Overall, performance improved by about 65\% on the TPC-H dataset and over 50\% on the Twitter dataset, further confirming the effectiveness of our proposed algorithm.

\subsubsection{Ablations on Clustering}
\begin{figure}
\centering
\includegraphics[width=1\linewidth]{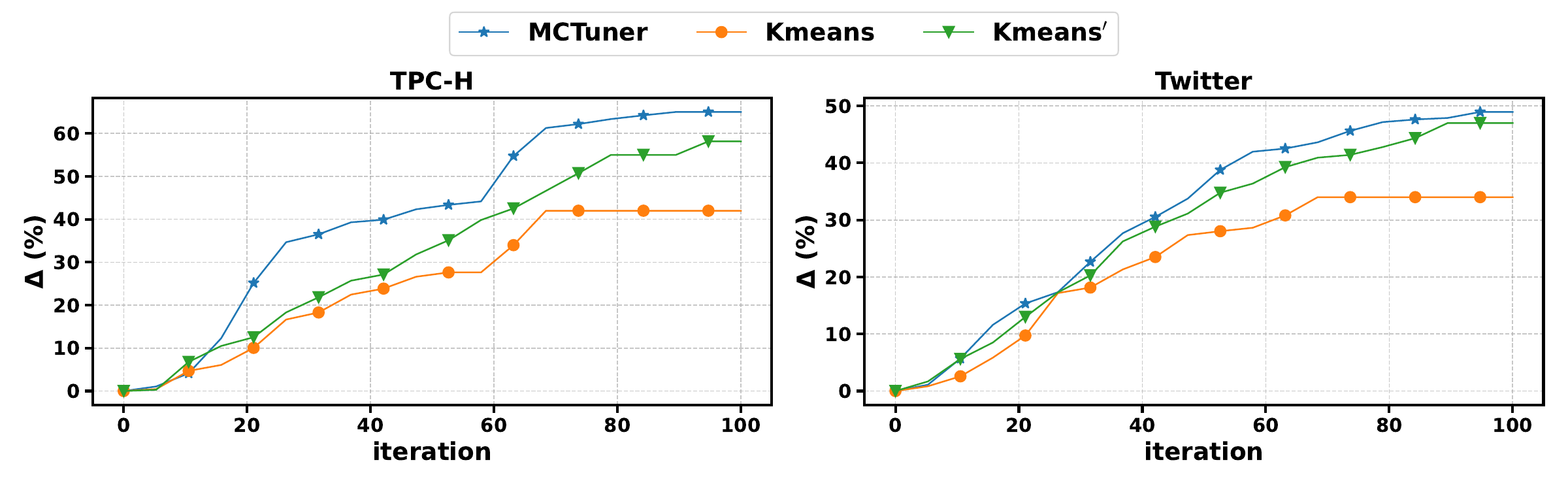}
\vspace{-2em}
\caption{Ablation study on the clustering method.}
\vspace{-2em}
\label{FIG_clustering}
\end{figure}

In the spatial decomposition algorithm, we propose a two-stage clustering method for splitting leaf nodes and compare it to the traditional K-Means algorithm. We found that K-Means often fails to continue classification, prematurely halting the tuning process (Fig.~\ref{FIG_clustering}). This issue likely stems from the use of the L2 norm, which may not accurately capture data point distributions. To address this, we replaced the L2 norm with cosine similarity (K-Means$^\prime$), eliminating the dimensionality reduction step used in \systemname, thus enabling smoother tuning.

During the early tuning stages, K-Means$^\prime$ identified better knob configurations faster than the original K-Means. Although its final performance was comparable to \systemname's, \systemname~achieved optimal configurations in fewer iterations. These results underscore the effectiveness of \systemname's two-stage clustering method.


\subsubsection{Component-wise Ablation of \textsc{MCTuner}}
\begin{figure}
    \centering
    \begin{minipage}{0.49\linewidth}
        \centering
        \includegraphics[width=1\linewidth]{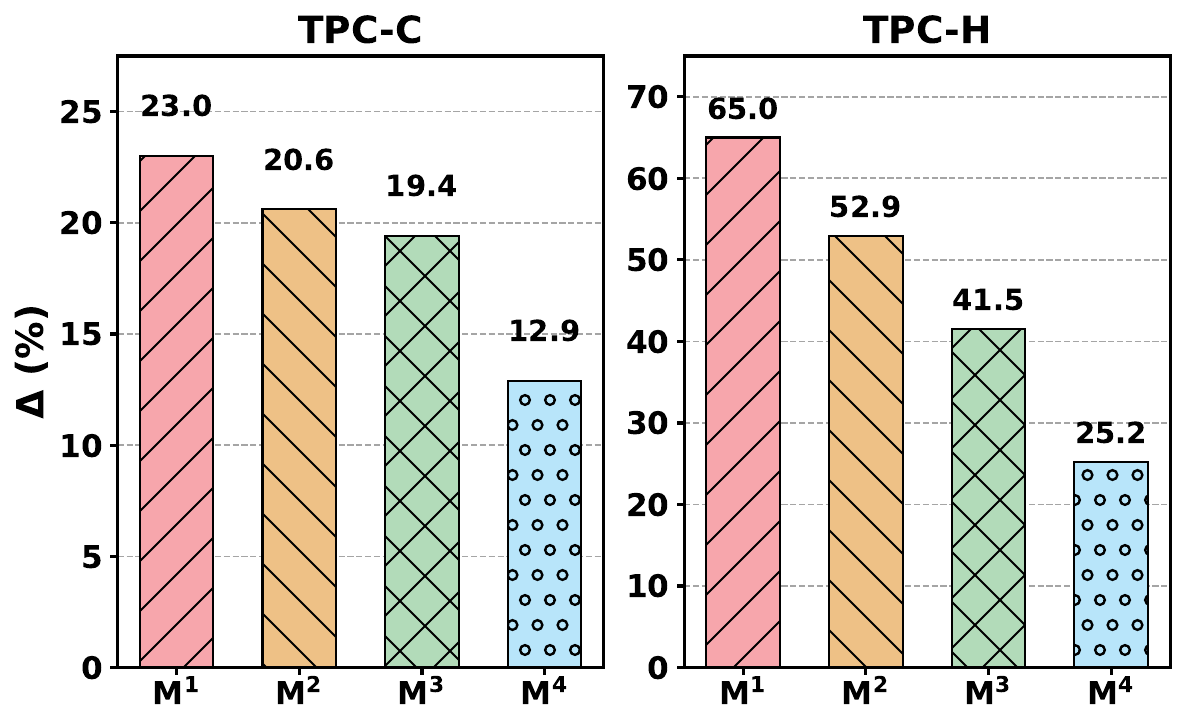}
        \vspace{-2em}
        \caption{Component-wise ablation of \systemname.}
        \label{ablation_space}
    \end{minipage}
    \hfill
    \begin{minipage}{0.49\linewidth}
        \centering
        \includegraphics[width=1\linewidth]{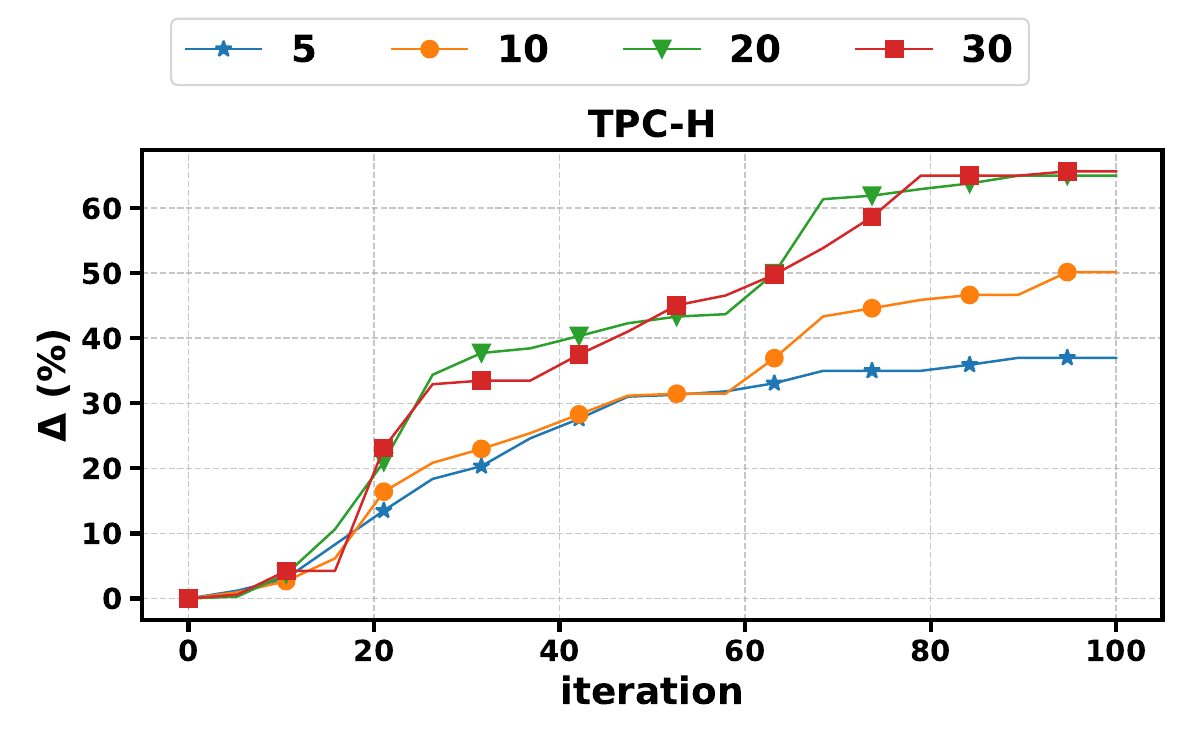}
        \vspace{-2em}
        \caption{Ablation study on knob number .}
        \label{ablation_knob}
    \end{minipage}
    \vspace{-1em}
\end{figure}

We conduct an ablation study to evaluate the contribution of each component within \systemname~to overall tuning performance. The results are presented in Figure\ref{ablation_space}, where M$^1$ represents the original \systemname. M$^2$ replaces the MoE-based knob selection and compression with a zero-shot LLM. M$^3$ removes the MoE mechanism entirely, and M$^4$ substitutes the spatial decomposition algorithm with BO.

The results show that each module of \systemname~contributes significantly to tuning efficiency and final performance on both the TPC-C and TPC-H workloads. The effect is particularly pronounced on the TPC-H dataset. Compared to the zero-shot LLM, the MoE module provides a 12.1 percent improvement in tuning performance, while the spatial decomposition algorithm leads to a substantial 39.8 percent gain. Similar trends are observed on the TPC-C dataset, further validating the effectiveness of each component.

These findings confirm the necessity of both modules and highlight their complementary roles. The MoE module facilitates knowledge-driven exploration, while the spatial decomposition mechanism improves local exploration efficiency and reduces redundant sampling.

\subsubsection[Ablations on Knob Number]{Ablations on Knob Number}
In Section~\ref{adaptive}, we customize the number of knobs selected, experimenting with four different counts: 5, 10, 20, and 30, as shown in Fig.~\ref{ablation_knob}. The results show that while increasing the number of knobs generally improves tuning performance, the improvement plateaus when moving from 20 to 30 knobs. This observation aligns with prior findings in \cite{kanellis2020too}, which suggest that only a small subset of knobs significantly impacts system performance. As the number of selected knobs grows, the configuration space expands exponentially, making the optimization process more costly and potentially less stable due to increased search complexity. On the other hand, selecting too few knobs may lead to the exclusion of influential parameters, limiting the achievable performance improvements. Our results demonstrate that selecting 20 knobs provides sufficient flexibility to capture key tuning opportunities while keeping the search space tractable. This balance enables more efficient exploration and stable optimization, validating \systemname's effectiveness in identifying a compact yet expressive set of knobs.



\section{Conclusion}
This paper presents \systemname, an adaptive database tuning framework that leverages LLM-guided knob selection and spatial decomposition to improve tuning efficiency. Extensive experiments are conducted to validate the effectiveness of \systemname, which consistently outperforms existing state-of-the-art methods across diverse workloads. Future work includes extending \systemname~to multi-database environments and enabling real-time adaptation to evolving workloads.


\clearpage
\bibliographystyle{ACM-Reference-Format}
\bibliography{ref}


\begin{thebibliography}{60}


\ifx \showCODEN    \undefined \def \showCODEN     #1{\unskip}     \fi
\ifx \showISBNx    \undefined \def \showISBNx     #1{\unskip}     \fi
\ifx \showISBNxiii \undefined \def \showISBNxiii  #1{\unskip}     \fi
\ifx \showISSN     \undefined \def \showISSN      #1{\unskip}     \fi
\ifx \showLCCN     \undefined \def \showLCCN      #1{\unskip}     \fi
\ifx \shownote     \undefined \def \shownote      #1{#1}          \fi
\ifx \showarticletitle \undefined \def \showarticletitle #1{#1}   \fi
\ifx \showURL      \undefined \def \showURL       {\relax}        \fi
\providecommand\bibfield[2]{#2}
\providecommand\bibinfo[2]{#2}
\providecommand\natexlab[1]{#1}
\providecommand\showeprint[2][]{arXiv:#2}

\bibitem[Ahmed et~al\mbox{.}(2020)]%
        {ahmed2020k}
\bibfield{author}{\bibinfo{person}{Mohiuddin Ahmed}, \bibinfo{person}{Raihan Seraj}, {and} \bibinfo{person}{Syed Mohammed~Shamsul Islam}.} \bibinfo{year}{2020}\natexlab{}.
\newblock \showarticletitle{The k-means algorithm: A comprehensive survey and performance evaluation}.
\newblock \bibinfo{journal}{\emph{Electronics}} \bibinfo{volume}{9}, \bibinfo{number}{8} (\bibinfo{year}{2020}), \bibinfo{pages}{1295}.
\newblock


\bibitem[Akioyamen et~al\mbox{.}(2024)]%
        {akioyamen2024unreasonable}
\bibfield{author}{\bibinfo{person}{Peter Akioyamen}, \bibinfo{person}{Zixuan Yi}, {and} \bibinfo{person}{Ryan Marcus}.} \bibinfo{year}{2024}\natexlab{}.
\newblock \showarticletitle{The Unreasonable Effectiveness of LLMs for Query Optimization}.
\newblock \bibinfo{journal}{\emph{arXiv preprint arXiv:2411.02862}} (\bibinfo{year}{2024}).
\newblock


\bibitem[Ansel et~al\mbox{.}(2014)]%
        {ansel2014opentuner}
\bibfield{author}{\bibinfo{person}{Jason Ansel}, \bibinfo{person}{Shoaib Kamil}, \bibinfo{person}{Kalyan Veeramachaneni}, \bibinfo{person}{Jonathan Ragan-Kelley}, \bibinfo{person}{Jeffrey Bosboom}, \bibinfo{person}{Una-May O'Reilly}, {and} \bibinfo{person}{Saman Amarasinghe}.} \bibinfo{year}{2014}\natexlab{}.
\newblock \showarticletitle{Opentuner: An extensible framework for program autotuning}. In \bibinfo{booktitle}{\emph{Proceedings of the 23rd international conference on Parallel architectures and compilation}}. \bibinfo{pages}{303--316}.
\newblock


\bibitem[Cai et~al\mbox{.}(2022)]%
        {cai2022hunter}
\bibfield{author}{\bibinfo{person}{Baoqing Cai}, \bibinfo{person}{Yu Liu}, \bibinfo{person}{Ce Zhang}, \bibinfo{person}{Guangyu Zhang}, \bibinfo{person}{Ke Zhou}, \bibinfo{person}{Li Liu}, \bibinfo{person}{Chunhua Li}, \bibinfo{person}{Bin Cheng}, \bibinfo{person}{Jie Yang}, {and} \bibinfo{person}{Jiashu Xing}.} \bibinfo{year}{2022}\natexlab{}.
\newblock \showarticletitle{HUNTER: an online cloud database hybrid tuning system for personalized requirements}. In \bibinfo{booktitle}{\emph{Proceedings of the 2022 International Conference on Management of Data}}. \bibinfo{pages}{646--659}.
\newblock


\bibitem[Cereda et~al\mbox{.}(2021)]%
        {cereda2021cGPTuner}
\bibfield{author}{\bibinfo{person}{Stefano Cereda}, \bibinfo{person}{Stefano Valladares}, \bibinfo{person}{Paolo Cremonesi}, \bibinfo{person}{Stefano Doni}, {et~al\mbox{.}}} \bibinfo{year}{2021}\natexlab{}.
\newblock \showarticletitle{Cgptuner: a contextual gaussian process bandit approach for the automatic tuning of it configurations under varying workload conditions}.
\newblock \bibinfo{journal}{\emph{Proceedings of the VLDB Endowment}} \bibinfo{volume}{14}, \bibinfo{number}{8} (\bibinfo{year}{2021}), \bibinfo{pages}{1401--1413}.
\newblock


\bibitem[Dageville and Zait(2002)]%
        {dageville2002sql}
\bibfield{author}{\bibinfo{person}{Beno{\^\i}t Dageville} {and} \bibinfo{person}{Mohamed Zait}.} \bibinfo{year}{2002}\natexlab{}.
\newblock \showarticletitle{SQL memory management in Oracle9i}. In \bibinfo{booktitle}{\emph{VLDB'02: Proceedings of the 28th International Conference on Very Large Databases}}. Elsevier, \bibinfo{pages}{962--973}.
\newblock


\bibitem[Debnath et~al\mbox{.}(2008)]%
        {debnath2008sard}
\bibfield{author}{\bibinfo{person}{Biplob~K Debnath}, \bibinfo{person}{David~J Lilja}, {and} \bibinfo{person}{Mohamed~F Mokbel}.} \bibinfo{year}{2008}\natexlab{}.
\newblock \showarticletitle{SARD: A statistical approach for ranking database tuning parameters}. In \bibinfo{booktitle}{\emph{2008 IEEE 24th International Conference on Data Engineering Workshop}}. IEEE, \bibinfo{pages}{11--18}.
\newblock


\bibitem[Difallah et~al\mbox{.}(2013)]%
        {DifallahPCC13}
\bibfield{author}{\bibinfo{person}{Djellel~Eddine Difallah}, \bibinfo{person}{Andrew Pavlo}, \bibinfo{person}{Carlo Curino}, {and} \bibinfo{person}{Philippe Cudr{\'e}-Mauroux}.} \bibinfo{year}{2013}\natexlab{}.
\newblock \showarticletitle{OLTP-Bench: An Extensible Testbed for Benchmarking Relational Databases}.
\newblock \bibinfo{journal}{\emph{PVLDB}} \bibinfo{volume}{7}, \bibinfo{number}{4} (\bibinfo{year}{2013}), \bibinfo{pages}{277--288}.
\newblock
\urldef\tempurl%
\url{http://www.vldb.org/pvldb/vol7/p277-difallah.pdf}
\showURL{%
\tempurl}


\bibitem[Duan et~al\mbox{.}(2009)]%
        {duan2009tuning}
\bibfield{author}{\bibinfo{person}{Songyun Duan}, \bibinfo{person}{Vamsidhar Thummala}, {and} \bibinfo{person}{Shivnath Babu}.} \bibinfo{year}{2009}\natexlab{}.
\newblock \showarticletitle{Tuning database configuration parameters with ituned}.
\newblock \bibinfo{journal}{\emph{Proceedings of the VLDB Endowment}} \bibinfo{volume}{2}, \bibinfo{number}{1} (\bibinfo{year}{2009}), \bibinfo{pages}{1246--1257}.
\newblock


\bibitem[Ge et~al\mbox{.}(2021)]%
        {ge2021watuning}
\bibfield{author}{\bibinfo{person}{Jia-Ke Ge}, \bibinfo{person}{Yan-Feng Chai}, {and} \bibinfo{person}{Yun-Peng Chai}.} \bibinfo{year}{2021}\natexlab{}.
\newblock \showarticletitle{WATuning: a workload-aware tuning system with attention-based deep reinforcement learning}.
\newblock \bibinfo{journal}{\emph{Journal of Computer Science and Technology}} \bibinfo{volume}{36}, \bibinfo{number}{4} (\bibinfo{year}{2021}), \bibinfo{pages}{741--761}.
\newblock


\bibitem[Giannakouris and Trummer(2024)]%
        {giannakouris2024dbg}
\bibfield{author}{\bibinfo{person}{Victor Giannakouris} {and} \bibinfo{person}{Immanuel Trummer}.} \bibinfo{year}{2024}\natexlab{}.
\newblock \showarticletitle{DBG-PT: A Large Language Model Assisted Query Performance Regression Debugger}.
\newblock \bibinfo{journal}{\emph{Proceedings of the VLDB Endowment}} \bibinfo{volume}{17}, \bibinfo{number}{12} (\bibinfo{year}{2024}), \bibinfo{pages}{4337--4340}.
\newblock


\bibitem[Giannakouris and Trummer(2025)]%
        {giannakouris2025lambda}
\bibfield{author}{\bibinfo{person}{Victor Giannakouris} {and} \bibinfo{person}{Immanuel Trummer}.} \bibinfo{year}{2025}\natexlab{}.
\newblock \showarticletitle{$\lambda$-Tune: Harnessing Large Language Models for Automated Database System Tuning}.
\newblock \bibinfo{journal}{\emph{Proceedings of the ACM on Management of Data}} \bibinfo{volume}{3}, \bibinfo{number}{1} (\bibinfo{year}{2025}), \bibinfo{pages}{1--26}.
\newblock


\bibitem[Graves(2011)]%
        {graves2011practical}
\bibfield{author}{\bibinfo{person}{Alex Graves}.} \bibinfo{year}{2011}\natexlab{}.
\newblock \showarticletitle{Practical variational inference for neural networks}.
\newblock \bibinfo{journal}{\emph{Advances in neural information processing systems}}  \bibinfo{volume}{24} (\bibinfo{year}{2011}).
\newblock


\bibitem[Hern{\'a}ndez-Lobato and Adams(2015)]%
        {hernandez2015probabilistic}
\bibfield{author}{\bibinfo{person}{Jos{\'e}~Miguel Hern{\'a}ndez-Lobato} {and} \bibinfo{person}{Ryan Adams}.} \bibinfo{year}{2015}\natexlab{}.
\newblock \showarticletitle{Probabilistic backpropagation for scalable learning of bayesian neural networks}. In \bibinfo{booktitle}{\emph{International conference on machine learning}}. PMLR, \bibinfo{pages}{1861--1869}.
\newblock


\bibitem[Hoffman et~al\mbox{.}(2011)]%
        {hoffman2011portfolio}
\bibfield{author}{\bibinfo{person}{Matthew Hoffman}, \bibinfo{person}{Eric Brochu}, \bibinfo{person}{Nando De~Freitas}, {et~al\mbox{.}}} \bibinfo{year}{2011}\natexlab{}.
\newblock \showarticletitle{Portfolio Allocation for Bayesian Optimization.}. In \bibinfo{booktitle}{\emph{UAI}}, Vol.~\bibinfo{volume}{11}. \bibinfo{pages}{327--336}.
\newblock


\bibitem[Huang et~al\mbox{.}(2025)]%
        {huang2025e2etuneendtoendknobtuning}
\bibfield{author}{\bibinfo{person}{Xinmei Huang}, \bibinfo{person}{Haoyang Li}, \bibinfo{person}{Jing Zhang}, \bibinfo{person}{Xinxin Zhao}, \bibinfo{person}{Zhiming Yao}, \bibinfo{person}{Yiyan Li}, \bibinfo{person}{Tieying Zhang}, \bibinfo{person}{Jianjun Chen}, \bibinfo{person}{Hong Chen}, {and} \bibinfo{person}{Cuiping Li}.} \bibinfo{year}{2025}\natexlab{}.
\newblock \bibinfo{title}{E2ETune: End-to-End Knob Tuning via Fine-tuned Generative Language Model}.
\newblock
\showeprint[arxiv]{2404.11581}~[cs.AI]
\urldef\tempurl%
\url{https://arxiv.org/abs/2404.11581}
\showURL{%
\tempurl}


\bibitem[Hutter et~al\mbox{.}(2011)]%
        {hutter2011sequential}
\bibfield{author}{\bibinfo{person}{Frank Hutter}, \bibinfo{person}{Holger~H Hoos}, {and} \bibinfo{person}{Kevin Leyton-Brown}.} \bibinfo{year}{2011}\natexlab{}.
\newblock \showarticletitle{Sequential model-based optimization for general algorithm configuration}. In \bibinfo{booktitle}{\emph{Learning and intelligent optimization: 5th international conference, LION 5, rome, Italy, January 17-21, 2011. selected papers 5}}. Springer, \bibinfo{pages}{507--523}.
\newblock


\bibitem[Kanellis et~al\mbox{.}(2020)]%
        {kanellis2020too}
\bibfield{author}{\bibinfo{person}{Konstantinos Kanellis}, \bibinfo{person}{Ramnatthan Alagappan}, {and} \bibinfo{person}{Shivaram Venkataraman}.} \bibinfo{year}{2020}\natexlab{}.
\newblock \showarticletitle{Too many knobs to tune? towards faster database tuning by pre-selecting important knobs}. In \bibinfo{booktitle}{\emph{12th USENIX Workshop on Hot Topics in Storage and File Systems (HotStorage 20)}}.
\newblock


\bibitem[Kanellis et~al\mbox{.}(2022)]%
        {kanellis2022llamatune}
\bibfield{author}{\bibinfo{person}{Konstantinos Kanellis}, \bibinfo{person}{Cong Ding}, \bibinfo{person}{Brian Kroth}, \bibinfo{person}{Andreas M{\"u}ller}, \bibinfo{person}{Carlo Curino}, {and} \bibinfo{person}{Shivaram Venkataraman}.} \bibinfo{year}{2022}\natexlab{}.
\newblock \showarticletitle{LlamaTune: sample-efficient DBMS configuration tuning}.
\newblock \bibinfo{journal}{\emph{arXiv preprint arXiv:2203.05128}} (\bibinfo{year}{2022}).
\newblock


\bibitem[Kunjir and Babu(2020)]%
        {kunjir2020black}
\bibfield{author}{\bibinfo{person}{Mayuresh Kunjir} {and} \bibinfo{person}{Shivnath Babu}.} \bibinfo{year}{2020}\natexlab{}.
\newblock \showarticletitle{Black or white? how to develop an autotuner for memory-based analytics}. In \bibinfo{booktitle}{\emph{Proceedings of the 2020 ACM SIGMOD International Conference on Management of Data}}. \bibinfo{pages}{1667--1683}.
\newblock


\bibitem[Lao et~al\mbox{.}(2023)]%
        {lao2023GPTuner}
\bibfield{author}{\bibinfo{person}{Jiale Lao}, \bibinfo{person}{Yibo Wang}, \bibinfo{person}{Yufei Li}, \bibinfo{person}{Jianping Wang}, \bibinfo{person}{Yunjia Zhang}, \bibinfo{person}{Zhiyuan Cheng}, \bibinfo{person}{Wanghu Chen}, \bibinfo{person}{Mingjie Tang}, {and} \bibinfo{person}{Jianguo Wang}.} \bibinfo{year}{2023}\natexlab{}.
\newblock \showarticletitle{Gptuner: A manual-reading database tuning system via gpt-guided bayesian optimization}.
\newblock \bibinfo{journal}{\emph{arXiv preprint arXiv:2311.03157}} (\bibinfo{year}{2023}).
\newblock


\bibitem[Leis et~al\mbox{.}(2015)]%
        {leis2015good}
\bibfield{author}{\bibinfo{person}{Viktor Leis}, \bibinfo{person}{Andrey Gubichev}, \bibinfo{person}{Atanas Mirchev}, \bibinfo{person}{Peter Boncz}, \bibinfo{person}{Alfons Kemper}, {and} \bibinfo{person}{Thomas Neumann}.} \bibinfo{year}{2015}\natexlab{}.
\newblock \showarticletitle{How good are query optimizers, really?}
\newblock \bibinfo{journal}{\emph{Proceedings of the VLDB Endowment}} \bibinfo{volume}{9}, \bibinfo{number}{3} (\bibinfo{year}{2015}), \bibinfo{pages}{204--215}.
\newblock


\bibitem[Li et~al\mbox{.}(2024b)]%
        {li2024dawn}
\bibfield{author}{\bibinfo{person}{Boyan Li}, \bibinfo{person}{Yuyu Luo}, \bibinfo{person}{Chengliang Chai}, \bibinfo{person}{Guoliang Li}, {and} \bibinfo{person}{Nan Tang}.} \bibinfo{year}{2024}\natexlab{b}.
\newblock \showarticletitle{The dawn of natural language to SQL: are we fully ready?}
\newblock \bibinfo{journal}{\emph{arXiv preprint arXiv:2406.01265}} (\bibinfo{year}{2024}).
\newblock


\bibitem[Li et~al\mbox{.}(2019)]%
        {li2019qtune}
\bibfield{author}{\bibinfo{person}{Guoliang Li}, \bibinfo{person}{Xuanhe Zhou}, \bibinfo{person}{Shifu Li}, {and} \bibinfo{person}{Bo Gao}.} \bibinfo{year}{2019}\natexlab{}.
\newblock \showarticletitle{Qtune: A query-aware database tuning system with deep reinforcement learning}.
\newblock \bibinfo{journal}{\emph{Proceedings of the VLDB Endowment}} \bibinfo{volume}{12}, \bibinfo{number}{12} (\bibinfo{year}{2019}), \bibinfo{pages}{2118--2130}.
\newblock


\bibitem[Li et~al\mbox{.}(2023)]%
        {li2023resdsql}
\bibfield{author}{\bibinfo{person}{Haoyang Li}, \bibinfo{person}{Jing Zhang}, \bibinfo{person}{Cuiping Li}, {and} \bibinfo{person}{Hong Chen}.} \bibinfo{year}{2023}\natexlab{}.
\newblock \showarticletitle{Resdsql: Decoupling schema linking and skeleton parsing for text-to-sql}. In \bibinfo{booktitle}{\emph{Proceedings of the AAAI Conference on Artificial Intelligence}}, Vol.~\bibinfo{volume}{37}. \bibinfo{pages}{13067--13075}.
\newblock


\bibitem[Li et~al\mbox{.}(2024d)]%
        {li2024codes}
\bibfield{author}{\bibinfo{person}{Haoyang Li}, \bibinfo{person}{Jing Zhang}, \bibinfo{person}{Hanbing Liu}, \bibinfo{person}{Ju Fan}, \bibinfo{person}{Xiaokang Zhang}, \bibinfo{person}{Jun Zhu}, \bibinfo{person}{Renjie Wei}, \bibinfo{person}{Hongyan Pan}, \bibinfo{person}{Cuiping Li}, {and} \bibinfo{person}{Hong Chen}.} \bibinfo{year}{2024}\natexlab{d}.
\newblock \showarticletitle{Codes: Towards building open-source language models for text-to-sql}.
\newblock \bibinfo{journal}{\emph{Proceedings of the ACM on Management of Data}} \bibinfo{volume}{2}, \bibinfo{number}{3} (\bibinfo{year}{2024}), \bibinfo{pages}{1--28}.
\newblock


\bibitem[Li et~al\mbox{.}(2024a)]%
        {li2024large}
\bibfield{author}{\bibinfo{person}{Yiyan Li}, \bibinfo{person}{Haoyang Li}, \bibinfo{person}{Zhao Pu}, \bibinfo{person}{Jing Zhang}, \bibinfo{person}{Xinyi Zhang}, \bibinfo{person}{Tao Ji}, \bibinfo{person}{Luming Sun}, \bibinfo{person}{Cuiping Li}, {and} \bibinfo{person}{Hong Chen}.} \bibinfo{year}{2024}\natexlab{a}.
\newblock \showarticletitle{Is Large Language Model Good at Database Knob Tuning? A Comprehensive Experimental Evaluation}.
\newblock \bibinfo{journal}{\emph{arXiv preprint arXiv:2408.02213}} (\bibinfo{year}{2024}).
\newblock


\bibitem[Li et~al\mbox{.}(2024c)]%
        {li2024llm}
\bibfield{author}{\bibinfo{person}{Zhaodonghui Li}, \bibinfo{person}{Haitao Yuan}, \bibinfo{person}{Huiming Wang}, \bibinfo{person}{Gao Cong}, {and} \bibinfo{person}{Lidong Bing}.} \bibinfo{year}{2024}\natexlab{c}.
\newblock \showarticletitle{LLM-R2: A Large Language Model Enhanced Rule-based Rewrite System for Boosting Query Efficiency}.
\newblock \bibinfo{journal}{\emph{arXiv preprint arXiv:2404.12872}} (\bibinfo{year}{2024}).
\newblock


\bibitem[McKay(1992)]%
        {mckay1992latin}
\bibfield{author}{\bibinfo{person}{Michael~D McKay}.} \bibinfo{year}{1992}\natexlab{}.
\newblock \showarticletitle{Latin hypercube sampling as a tool in uncertainty analysis of computer models}. In \bibinfo{booktitle}{\emph{Proceedings of the 24th conference on Winter simulation}}. \bibinfo{pages}{557--564}.
\newblock


\bibitem[Ouyang et~al\mbox{.}(2025)]%
        {ouyang2025rcrank}
\bibfield{author}{\bibinfo{person}{Biao Ouyang}, \bibinfo{person}{Yingying Zhang}, \bibinfo{person}{Hanyin Cheng}, \bibinfo{person}{Yang Shu}, \bibinfo{person}{Chenjuan Guo}, \bibinfo{person}{Bin Yang}, \bibinfo{person}{Qingsong Wen}, \bibinfo{person}{Lunting Fan}, {and} \bibinfo{person}{Christian~S Jensen}.} \bibinfo{year}{2025}\natexlab{}.
\newblock \showarticletitle{RCRank: Multimodal Ranking of Root Causes of Slow Queries in Cloud Database Systems}.
\newblock \bibinfo{journal}{\emph{arXiv preprint arXiv:2503.04252}} (\bibinfo{year}{2025}).
\newblock


\bibitem[Park and Jun(2009)]%
        {park2009simple}
\bibfield{author}{\bibinfo{person}{Hae-Sang Park} {and} \bibinfo{person}{Chi-Hyuck Jun}.} \bibinfo{year}{2009}\natexlab{}.
\newblock \showarticletitle{A simple and fast algorithm for K-medoids clustering}.
\newblock \bibinfo{journal}{\emph{Expert systems with applications}} \bibinfo{volume}{36}, \bibinfo{number}{2} (\bibinfo{year}{2009}), \bibinfo{pages}{3336--3341}.
\newblock


\bibitem[Pourreza and Rafiei(2023)]%
        {pourreza2023din}
\bibfield{author}{\bibinfo{person}{Mohammadreza Pourreza} {and} \bibinfo{person}{Davood Rafiei}.} \bibinfo{year}{2023}\natexlab{}.
\newblock \showarticletitle{Din-sql: Decomposed in-context learning of text-to-sql with self-correction}.
\newblock \bibinfo{journal}{\emph{Advances in Neural Information Processing Systems}}  \bibinfo{volume}{36} (\bibinfo{year}{2023}), \bibinfo{pages}{36339--36348}.
\newblock


\bibitem[Ren et~al\mbox{.}(2024)]%
        {ren2024purple}
\bibfield{author}{\bibinfo{person}{Tonghui Ren}, \bibinfo{person}{Yuankai Fan}, \bibinfo{person}{Zhenying He}, \bibinfo{person}{Ren Huang}, \bibinfo{person}{Jiaqi Dai}, \bibinfo{person}{Can Huang}, \bibinfo{person}{Yinan Jing}, \bibinfo{person}{Kai Zhang}, \bibinfo{person}{Yifan Yang}, {and} \bibinfo{person}{X~Sean Wang}.} \bibinfo{year}{2024}\natexlab{}.
\newblock \showarticletitle{Purple: Making a large language model a better sql writer}. In \bibinfo{booktitle}{\emph{2024 IEEE 40th International Conference on Data Engineering (ICDE)}}. IEEE, \bibinfo{pages}{15--28}.
\newblock


\bibitem[Sch{\"o}lkopf et~al\mbox{.}(1997)]%
        {scholkopf1997kernel}
\bibfield{author}{\bibinfo{person}{Bernhard Sch{\"o}lkopf}, \bibinfo{person}{Alexander Smola}, {and} \bibinfo{person}{Klaus-Robert M{\"u}ller}.} \bibinfo{year}{1997}\natexlab{}.
\newblock \showarticletitle{Kernel principal component analysis}. In \bibinfo{booktitle}{\emph{International conference on artificial neural networks}}. Springer, \bibinfo{pages}{583--588}.
\newblock


\bibitem[Silver et~al\mbox{.}(2014)]%
        {silver2014deterministic}
\bibfield{author}{\bibinfo{person}{David Silver}, \bibinfo{person}{Guy Lever}, \bibinfo{person}{Nicolas Heess}, \bibinfo{person}{Thomas Degris}, \bibinfo{person}{Daan Wierstra}, {and} \bibinfo{person}{Martin Riedmiller}.} \bibinfo{year}{2014}\natexlab{}.
\newblock \showarticletitle{Deterministic policy gradient algorithms}. In \bibinfo{booktitle}{\emph{International conference on machine learning}}. Pmlr, \bibinfo{pages}{387--395}.
\newblock


\bibitem[Singh et~al\mbox{.}(2024)]%
        {singh2024panda}
\bibfield{author}{\bibinfo{person}{Vikramank Singh}, \bibinfo{person}{Kapil~Eknath Vaidya}, \bibinfo{person}{Vinayshekhar~Bannihatti Kumar}, \bibinfo{person}{Sopan Khosla}, \bibinfo{person}{Murali Narayanaswamy}, \bibinfo{person}{Rashmi Gangadharaiah}, {and} \bibinfo{person}{Tim Kraska}.} \bibinfo{year}{2024}\natexlab{}.
\newblock \showarticletitle{Panda: Performance debugging for databases using LLM agents}.
\newblock  (\bibinfo{year}{2024}).
\newblock


\bibitem[Sivasubramaniam et~al\mbox{.}(2024)]%
        {sivasubramaniam2024sm3}
\bibfield{author}{\bibinfo{person}{Sithursan Sivasubramaniam}, \bibinfo{person}{Cedric~E Osei-Akoto}, \bibinfo{person}{Yi Zhang}, \bibinfo{person}{Kurt Stockinger}, {and} \bibinfo{person}{Jonathan Fuerst}.} \bibinfo{year}{2024}\natexlab{}.
\newblock \showarticletitle{SM3-Text-to-Query: Synthetic Multi-Model Medical Text-to-Query Benchmark}.
\newblock \bibinfo{journal}{\emph{Advances in Neural Information Processing Systems}}  \bibinfo{volume}{37} (\bibinfo{year}{2024}), \bibinfo{pages}{88627--88663}.
\newblock


\bibitem[Sullivan et~al\mbox{.}(2004)]%
        {sullivan2004using}
\bibfield{author}{\bibinfo{person}{David~G Sullivan}, \bibinfo{person}{Margo~I Seltzer}, {and} \bibinfo{person}{Avi Pfeffer}.} \bibinfo{year}{2004}\natexlab{}.
\newblock \showarticletitle{Using probabilistic reasoning to automate software tuning}.
\newblock \bibinfo{journal}{\emph{ACM SIGMETRICS Performance Evaluation Review}} \bibinfo{volume}{32}, \bibinfo{number}{1} (\bibinfo{year}{2004}), \bibinfo{pages}{404--405}.
\newblock


\bibitem[Sun et~al\mbox{.}(2024)]%
        {sun2024r}
\bibfield{author}{\bibinfo{person}{Zhaoyan Sun}, \bibinfo{person}{Xuanhe Zhou}, {and} \bibinfo{person}{Guoliang Li}.} \bibinfo{year}{2024}\natexlab{}.
\newblock \showarticletitle{R-Bot: An LLM-based Query Rewrite System}.
\newblock \bibinfo{journal}{\emph{arXiv preprint arXiv:2412.01661}} (\bibinfo{year}{2024}).
\newblock


\bibitem[Trummer(2022a)]%
        {trummer2022codexdb}
\bibfield{author}{\bibinfo{person}{Immanuel Trummer}.} \bibinfo{year}{2022}\natexlab{a}.
\newblock \showarticletitle{CodexDB: Synthesizing code for query processing from natural language instructions using GPT-3 Codex}.
\newblock \bibinfo{journal}{\emph{Proceedings of the VLDB Endowment}} \bibinfo{volume}{15}, \bibinfo{number}{11} (\bibinfo{year}{2022}), \bibinfo{pages}{2921--2928}.
\newblock


\bibitem[Trummer(2022b)]%
        {trummer2022db}
\bibfield{author}{\bibinfo{person}{Immanuel Trummer}.} \bibinfo{year}{2022}\natexlab{b}.
\newblock \showarticletitle{DB-BERT: a Database Tuning Tool that" Reads the Manual"}. In \bibinfo{booktitle}{\emph{Proceedings of the 2022 international conference on management of data}}. \bibinfo{pages}{190--203}.
\newblock


\bibitem[Van~Aken et~al\mbox{.}(2017)]%
        {van2017automatic}
\bibfield{author}{\bibinfo{person}{Dana Van~Aken}, \bibinfo{person}{Andrew Pavlo}, \bibinfo{person}{Geoffrey~J Gordon}, {and} \bibinfo{person}{Bohan Zhang}.} \bibinfo{year}{2017}\natexlab{}.
\newblock \showarticletitle{Automatic database management system tuning through large-scale machine learning}. In \bibinfo{booktitle}{\emph{Proceedings of the 2017 ACM international conference on management of data}}. \bibinfo{pages}{1009--1024}.
\newblock


\bibitem[Van~Aken et~al\mbox{.}(2021)]%
        {van2021inquiry}
\bibfield{author}{\bibinfo{person}{Dana Van~Aken}, \bibinfo{person}{Dongsheng Yang}, \bibinfo{person}{Sebastien Brillard}, \bibinfo{person}{Ari Fiorino}, \bibinfo{person}{Bohan Zhang}, \bibinfo{person}{Christian Bilien}, {and} \bibinfo{person}{Andrew Pavlo}.} \bibinfo{year}{2021}\natexlab{}.
\newblock \showarticletitle{An inquiry into machine learning-based automatic configuration tuning services on real-world database management systems}.
\newblock \bibinfo{journal}{\emph{Proceedings of the VLDB Endowment}} \bibinfo{volume}{14}, \bibinfo{number}{7} (\bibinfo{year}{2021}), \bibinfo{pages}{1241--1253}.
\newblock


\bibitem[Von~Luxburg(2007)]%
        {von2007tutorial}
\bibfield{author}{\bibinfo{person}{Ulrike Von~Luxburg}.} \bibinfo{year}{2007}\natexlab{}.
\newblock \showarticletitle{A tutorial on spectral clustering}.
\newblock \bibinfo{journal}{\emph{Statistics and computing}}  \bibinfo{volume}{17} (\bibinfo{year}{2007}), \bibinfo{pages}{395--416}.
\newblock


\bibitem[Wang et~al\mbox{.}(2021)]%
        {wang2021udo}
\bibfield{author}{\bibinfo{person}{Junxiong Wang}, \bibinfo{person}{Immanuel Trummer}, {and} \bibinfo{person}{Debabrota Basu}.} \bibinfo{year}{2021}\natexlab{}.
\newblock \showarticletitle{UDO: universal database optimization using reinforcement learning}.
\newblock \bibinfo{journal}{\emph{arXiv preprint arXiv:2104.01744}} (\bibinfo{year}{2021}).
\newblock


\bibitem[Wang et~al\mbox{.}(2020)]%
        {wang2020learning}
\bibfield{author}{\bibinfo{person}{Linnan Wang}, \bibinfo{person}{Rodrigo Fonseca}, {and} \bibinfo{person}{Yuandong Tian}.} \bibinfo{year}{2020}\natexlab{}.
\newblock \showarticletitle{Learning search space partition for black-box optimization using monte carlo tree search}.
\newblock \bibinfo{journal}{\emph{Advances in Neural Information Processing Systems}}  \bibinfo{volume}{33} (\bibinfo{year}{2020}), \bibinfo{pages}{19511--19522}.
\newblock


\bibitem[Xiu et~al\mbox{.}(2024)]%
        {xiu2024query}
\bibfield{author}{\bibinfo{person}{Haibo Xiu}, \bibinfo{person}{Li Zhang}, \bibinfo{person}{Tieying Zhang}, \bibinfo{person}{Jun Yang}, {and} \bibinfo{person}{Jianjun Chen}.} \bibinfo{year}{2024}\natexlab{}.
\newblock \showarticletitle{Query Performance Explanation through Large Language Model for HTAP Systems}.
\newblock \bibinfo{journal}{\emph{arXiv preprint arXiv:2412.01709}} (\bibinfo{year}{2024}).
\newblock


\bibitem[Xu et~al\mbox{.}(2025)]%
        {xu2025chain}
\bibfield{author}{\bibinfo{person}{Silei Xu}, \bibinfo{person}{Wenhao Xie}, \bibinfo{person}{Lingxiao Zhao}, {and} \bibinfo{person}{Pengcheng He}.} \bibinfo{year}{2025}\natexlab{}.
\newblock \showarticletitle{Chain of draft: Thinking faster by writing less}.
\newblock \bibinfo{journal}{\emph{arXiv preprint arXiv:2502.18600}} (\bibinfo{year}{2025}).
\newblock


\bibitem[Zhang et~al\mbox{.}(2019)]%
        {zhang2019end}
\bibfield{author}{\bibinfo{person}{Ji Zhang}, \bibinfo{person}{Yu Liu}, \bibinfo{person}{Ke Zhou}, \bibinfo{person}{Guoliang Li}, \bibinfo{person}{Zhili Xiao}, \bibinfo{person}{Bin Cheng}, \bibinfo{person}{Jiashu Xing}, \bibinfo{person}{Yangtao Wang}, \bibinfo{person}{Tianheng Cheng}, \bibinfo{person}{Li Liu}, {et~al\mbox{.}}} \bibinfo{year}{2019}\natexlab{}.
\newblock \showarticletitle{An end-to-end automatic cloud database tuning system using deep reinforcement learning}. In \bibinfo{booktitle}{\emph{Proceedings of the 2019 international conference on management of data}}. \bibinfo{pages}{415--432}.
\newblock


\bibitem[Zhang and Babar(2024)]%
        {zhang2024automatic}
\bibfield{author}{\bibinfo{person}{Limeng Zhang} {and} \bibinfo{person}{M~Ali Babar}.} \bibinfo{year}{2024}\natexlab{}.
\newblock \showarticletitle{Automatic configuration tuning on cloud database: A survey}.
\newblock \bibinfo{journal}{\emph{arXiv preprint arXiv:2404.06043}} (\bibinfo{year}{2024}).
\newblock


\bibitem[Zhang et~al\mbox{.}(2021a)]%
        {zhang2021facilitating}
\bibfield{author}{\bibinfo{person}{Xinyi Zhang}, \bibinfo{person}{Zhuo Chang}, \bibinfo{person}{Yang Li}, \bibinfo{person}{Hong Wu}, \bibinfo{person}{Jian Tan}, \bibinfo{person}{Feifei Li}, {and} \bibinfo{person}{Bin Cui}.} \bibinfo{year}{2021}\natexlab{a}.
\newblock \showarticletitle{Facilitating database tuning with hyper-parameter optimization: a comprehensive experimental evaluation}.
\newblock \bibinfo{journal}{\emph{arXiv preprint arXiv:2110.12654}} (\bibinfo{year}{2021}).
\newblock


\bibitem[Zhang et~al\mbox{.}(2021b)]%
        {zhang2021restune}
\bibfield{author}{\bibinfo{person}{Xinyi Zhang}, \bibinfo{person}{Hong Wu}, \bibinfo{person}{Zhuo Chang}, \bibinfo{person}{Shuowei Jin}, \bibinfo{person}{Jian Tan}, \bibinfo{person}{Feifei Li}, \bibinfo{person}{Tieying Zhang}, {and} \bibinfo{person}{Bin Cui}.} \bibinfo{year}{2021}\natexlab{b}.
\newblock \showarticletitle{Restune: Resource oriented tuning boosted by meta-learning for cloud databases}. In \bibinfo{booktitle}{\emph{Proceedings of the 2021 international conference on management of data}}. \bibinfo{pages}{2102--2114}.
\newblock


\bibitem[Zhang et~al\mbox{.}(2022)]%
        {zhang2022towards}
\bibfield{author}{\bibinfo{person}{Xinyi Zhang}, \bibinfo{person}{Hong Wu}, \bibinfo{person}{Yang Li}, \bibinfo{person}{Jian Tan}, \bibinfo{person}{Feifei Li}, {and} \bibinfo{person}{Bin Cui}.} \bibinfo{year}{2022}\natexlab{}.
\newblock \showarticletitle{Towards dynamic and safe configuration tuning for cloud databases}. In \bibinfo{booktitle}{\emph{Proceedings of the 2022 International Conference on Management of Data}}. \bibinfo{pages}{631--645}.
\newblock


\bibitem[Zhang et~al\mbox{.}(2023a)]%
        {zhang2023efficient}
\bibfield{author}{\bibinfo{person}{Xinyi Zhang}, \bibinfo{person}{Hong Wu}, \bibinfo{person}{Yang Li}, \bibinfo{person}{Zhengju Tang}, \bibinfo{person}{Jian Tan}, \bibinfo{person}{Feifei Li}, {and} \bibinfo{person}{Bin Cui}.} \bibinfo{year}{2023}\natexlab{a}.
\newblock \showarticletitle{An efficient transfer learning based configuration adviser for database tuning}.
\newblock \bibinfo{journal}{\emph{Proceedings of the VLDB Endowment}} \bibinfo{volume}{17}, \bibinfo{number}{3} (\bibinfo{year}{2023}), \bibinfo{pages}{539--552}.
\newblock


\bibitem[Zhang et~al\mbox{.}(2023b)]%
        {zhang2023igniting}
\bibfield{author}{\bibinfo{person}{Zhuosheng Zhang}, \bibinfo{person}{Yao Yao}, \bibinfo{person}{Aston Zhang}, \bibinfo{person}{Xiangru Tang}, \bibinfo{person}{Xinbei Ma}, \bibinfo{person}{Zhiwei He}, \bibinfo{person}{Yiming Wang}, \bibinfo{person}{Mark Gerstein}, \bibinfo{person}{Rui Wang}, \bibinfo{person}{Gongshen Liu}, {et~al\mbox{.}}} \bibinfo{year}{2023}\natexlab{b}.
\newblock \showarticletitle{Igniting language intelligence: The hitchhiker's guide from chain-of-thought reasoning to language agents}.
\newblock \bibinfo{journal}{\emph{Comput. Surveys}} (\bibinfo{year}{2023}).
\newblock


\bibitem[Zhao et~al\mbox{.}(2025)]%
        {zhao2025llmidxadvis}
\bibfield{author}{\bibinfo{person}{Xinxin Zhao}, \bibinfo{person}{Haoyang Li}, \bibinfo{person}{Jing Zhang}, \bibinfo{person}{Xinmei Huang}, \bibinfo{person}{Tieying Zhang}, \bibinfo{person}{Jianjun Chen}, \bibinfo{person}{Rui Shi}, \bibinfo{person}{Cuiping Li}, {and} \bibinfo{person}{Hong Chen}.} \bibinfo{year}{2025}\natexlab{}.
\newblock \showarticletitle{LLMIdxAdvis: Resource-Efficient Index Advisor Utilizing Large Language Model}.
\newblock \bibinfo{journal}{\emph{arXiv preprint arXiv:2503.07884}} (\bibinfo{year}{2025}).
\newblock


\bibitem[Zhao et~al\mbox{.}(2023)]%
        {zhao2023automatic}
\bibfield{author}{\bibinfo{person}{Xinyang Zhao}, \bibinfo{person}{Xuanhe Zhou}, {and} \bibinfo{person}{Guoliang Li}.} \bibinfo{year}{2023}\natexlab{}.
\newblock \showarticletitle{Automatic database knob tuning: A survey}.
\newblock \bibinfo{journal}{\emph{IEEE Transactions on Knowledge and Data Engineering}} \bibinfo{volume}{35}, \bibinfo{number}{12} (\bibinfo{year}{2023}), \bibinfo{pages}{12470--12490}.
\newblock


\bibitem[Zhou et~al\mbox{.}(2024)]%
        {zhou2024breaking}
\bibfield{author}{\bibinfo{person}{Wei Zhou}, \bibinfo{person}{Chen Lin}, \bibinfo{person}{Xuanhe Zhou}, {and} \bibinfo{person}{Guoliang Li}.} \bibinfo{year}{2024}\natexlab{}.
\newblock \showarticletitle{Breaking It Down: An In-Depth Study of Index Advisors}.
\newblock \bibinfo{journal}{\emph{Proceedings of the VLDB Endowment}} \bibinfo{volume}{17}, \bibinfo{number}{10} (\bibinfo{year}{2024}), \bibinfo{pages}{2405--2418}.
\newblock


\bibitem[Zhou et~al\mbox{.}(2023)]%
        {zhou2023d}
\bibfield{author}{\bibinfo{person}{Xuanhe Zhou}, \bibinfo{person}{Guoliang Li}, \bibinfo{person}{Zhaoyan Sun}, \bibinfo{person}{Zhiyuan Liu}, \bibinfo{person}{Weize Chen}, \bibinfo{person}{Jianming Wu}, \bibinfo{person}{Jiesi Liu}, \bibinfo{person}{Ruohang Feng}, {and} \bibinfo{person}{Guoyang Zeng}.} \bibinfo{year}{2023}\natexlab{}.
\newblock \showarticletitle{D-bot: Database diagnosis system using large language models}.
\newblock \bibinfo{journal}{\emph{arXiv preprint arXiv:2312.01454}} (\bibinfo{year}{2023}).
\newblock


\bibitem[Zhu et~al\mbox{.}(2017)]%
        {zhu2017bestconfig}
\bibfield{author}{\bibinfo{person}{Yuqing Zhu}, \bibinfo{person}{Jianxun Liu}, \bibinfo{person}{Mengying Guo}, \bibinfo{person}{Yungang Bao}, \bibinfo{person}{Wenlong Ma}, \bibinfo{person}{Zhuoyue Liu}, \bibinfo{person}{Kunpeng Song}, {and} \bibinfo{person}{Yingchun Yang}.} \bibinfo{year}{2017}\natexlab{}.
\newblock \showarticletitle{Bestconfig: tapping the performance potential of systems via automatic configuration tuning}. In \bibinfo{booktitle}{\emph{Proceedings of the 2017 symposium on cloud computing}}. \bibinfo{pages}{338--350}.
\newblock


\end{thebibliography}
\end{sloppypar}
\end{document}